\documentclass{aastex631}
\usepackage{amsmath}
\usepackage{float}
\usepackage{graphicx}
\usepackage{soul}

\begin{document}

\title{A Method to Measure Photometries of 
Moderately-Saturated UVOT Sources}

\author{Hao Zhou}
\affiliation{Key Laboratory of Dark Matter and Space Astronomy, Purple Mountain Observatory, \\
Chinese Academy of Sciences, Nanjing 210023, China}
\affiliation{School of Astronomy and Space Science, University of Science and Technology of China, \\
Hefei 230026, China}
\affiliation{INAF/Brera Astronomical Observatory, via Bianchi 46, I-23807 Merate (LC), Italy}

\author{Zhi-Ping Jin}
\affiliation{Key Laboratory of Dark Matter and Space Astronomy, Purple Mountain Observatory, \\
Chinese Academy of Sciences, Nanjing 210023, China}
\affiliation{School of Astronomy and Space Science, University of Science and Technology of China, \\
Hefei 230026, China}

\author{Stefano Covino}
\affiliation{INAF/Brera Astronomical Observatory, via Bianchi 46, I-23807 Merate (LC), Italy}

\author{Yi-Zhong Fan}
\affiliation{Key Laboratory of Dark Matter and Space Astronomy, Purple Mountain Observatory, \\
Chinese Academy of Sciences, Nanjing 210023, China}
\affiliation{School of Astronomy and Space Science, University of Science and Technology of China, \\
Hefei 230026, China}

\author{Da-Ming Wei}
\affiliation{Key Laboratory of Dark Matter and Space Astronomy, Purple Mountain Observatory, \\
Chinese Academy of Sciences, Nanjing 210023, China}
\affiliation{School of Astronomy and Space Science, University of Science and Technology of China, \\
Hefei 230026, China}

\correspondingauthor{Zhi-Ping Jin, Yi-Zhong Fan}
\email{jin@pmo.ac.cn, yzfan@pmo.ac.cn}

\begin{abstract}
For bright transients such as Gamma-Ray Bursts (GRBs), the Ultra-Violet/Optical Telescope (UVOT) operates under event mode at early phases, which records incident positions and arrival time for each photon. The event file is able to be screened into many exposures to study the early light curve of GRBs with a high time resolution, including in particular the rapid brightening of the UV/Optical emission. Such a goal, however, is hampered for some extremely bright GRBs by the saturation in UVOT event images.
For moderately saturated UVOT sources, in this work we develop the method proposed in \cite{2023NatAs.tmp..135J} to recover their photometries. The basic idea is to assume a stable point spread function (PSF) of UVOT images, for which the counts in the core region (i.e., an aperture of a radius of 5 arcsec) and the wing region (i.e., an annulus ranging from 15 arcsec to 25 arcsec) should be a constant and the intrinsic flux can be reliably inferred with data in the ring. We demonstrate that in a given band, a tight correlation does hold among the background-removed count rates in the core and the wing. 
With the new method, the bright limit of measuring range for UVOT V and B bands increases $\sim1.7$ mag, while only $\sim0.7$ mag for U band due to the lack of bright calibration sources. Systematic uncertainties are $\sim0.2$ mag for V, B and U bands.
\end{abstract}


\section{Introduction} \label{sec:intro}
Ultra-Violet/Optical Telescope (UVOT) onboard \textit{Swift} satellite is designed to capture the early light of afterglow of Gamma-Ray Bursts (GRB)  or other rapid transients \citep{2005SSRv..120...95R}.
Once the satellite has been settled, UVOT will take a 150\,s exposure under event mode to make a finding chart for possible transients, and the typical start time is $\sim$100\,s after Burst Alert Telescope (BAT) trigger time.

Since the finding chart is taken under event mode, one can get arrival time and incident positions on detector for each photon and screen\footnote{Split the evt file into sub exposures with user defined time bins or other criteria. The description of the raw evt file is ``unscreened evt file" in archive and the command in UVOT FTOOLS is ``screen", the article uses ``screen" instead of ``split" or other similar words.} the event file with user specified time bins to derive the early light curve of the afterglow, e.g. GRB 080319B \citep{2008Natur.455..183R, 2013MNRAS.436.1684P}, GRB 081203A \citep{2009MNRAS.395L..21K}, GRB 130427A \citep{2014Sci...343...48M} and GRB 220101A \citep{2022GCN.31351....1K,2023NatAs.tmp..135J}.
However, some afterglows are so bright that they saturate in UVOT's images, e.g. GRB 080319B and GRB 130427A.
Efforts have been made to measure photometries of highly saturated sources with readout streaks, and such a method had been applied to GRB 080319B and GRB 130427A to recover their early photometries \citep{2013MNRAS.436.1684P,2014Sci...343...48M}.
However, the readout streak method introduces very large uncertainties if the source is moderately or barely saturated.

In \cite{2023NatAs.tmp..135J}, we proposed to use the photons collected at the tail of the Point Spread Function (PSF) to infer the intrinsic emission of moderately saturated UVOT sources \cite[see e.g.][for a similar approach in the infrared band]{2022AJ....163...46S}.
The fundamental assumption is that the PSF remains stable, and there exists a constant ratio between photons received in the core region (a circular region with a radius of 5 arcsec for UVOT) and the wing region (an annulus ranging from 15 arcsec to 25 arcsec). Please refer to Figure \ref{fig:saturation_pattern} for an illustration depicting the definitions of the core and wing regions.
The method used in this study is referred to as the PSF method to distinguish it from the readout streak method discussed in existing literature. \cite{2023NatAs.tmp..135J} have just briefly presented how to correct observations in White and V bands.
In this work we extensively examine the corrections in UVOT V, B and U bands. For the even bluer and White bands, the corrections are challenging because of the lack of calibration sources/observations. 

In Section 2, the principle of UVOT detection method is briefly introduced and the coincidence loss correction is described in detail.
The principle and calibration of the PSF method is discussed in Section 3, where photometric zeropoints of V and B bands are calibrated with Tycho-2 and Gaia Synthetic Photometry Catalogue (GSPC) sources and U-band zeropoint is calibrated with GSPC sources.
In Section 4, there are simple demonstrations for the PSF method.
The procedure and the valid range for the PSF method are summaried in section 5.
In this article, photometry results are reported in AB system and the pixel scale for images is 0.502\ arcsec/pixel, unless stated otherwise.

\section{Brief introduction to the UVOT}
\subsection{Detection principle and saturation count rate}
The final physical imaging equipment of UVOT is a small CCD of 385$\times$288 physical pixels (256$\times$256 for scientific usage, corresponding to 17$\times$17\,arcmin$^2$) but with high readout speed.
Under event mode, every $\sim$11\,ms, the CCD reads out an image to analyze positions of incident photons. With optoelectronic devices, incident photons are converted into electronic clouds, which will illuminate a phosphor screen to create many new photons.
Photons created from the phosphor screen are collected by the CCD via fibers connected on the anode of phosphor screen, and the Full Width Half Maximum (FWHM) of photons collected by the CCD is $\sim$1 physical pixel \citep{2000MNRAS.312...83F}.
There is an algorithm to restore positions of incident photons with images obtained by the CCD, and the accuracy for centroiding can reach 1/8 physical pixel \citep{2005SSRv..120...95R}, hence the size of final image is 2048$\times$2048 pixels.
If positions of 2 or more incident photons are too close in one frame (e.g., distances between peaks of photon clusters collected by the CCD are less than $\sim1$ physical pixel), the centroiding algorithm only restores one photon, which induces the so called COIncidence loss \citep[COI, ][]{2000MNRAS.312...83F}.
For UVOT, most photons from point sources are concentrated within a circular region with a radius of 5\,arcsec (corresponding to 1.25 physical pixels), and this region is defined as the optimal photometric aperture as well as the COI region for point sources, where the count rate is used to calculate the COI factor for point sources \citep{2008MNRAS.383..627P}.

The saturation occurs when at least 1 incident photon falls within the point source COI region per frame, since the centroiding algorithm can only record 1 incident photon.
In principle, the typical COI dominated region is a circular region with a radius of $\sim$2.25 physical pixels.
While in practice, moderately saturated sources exhibit a square pattern on source with a typical side length of $\sim20$\,arcsec (corresponding to a 5$\times$5 adjacent physical pixels), where the count rate in the region between the point souce COI region and the square is nearly 0\,count/s, and this square region represents the real COI dominated region of moderately saturated sources.
Nevertheless, the region dominated by the COI expands as the degree of saturation increases. 
Figure \ref{fig:saturation_pattern} shows saturation patterns for moderately and strongly saturated sources.

When the raw total count rate in the COI region is $>$10\,count/s for point sources, the coincidence loss is not negligible. The saturation limit of UVOT is 1 count per frame ($\sim$90\,count/s), since the coincidence loss makes any count rate greater than 1 count per frame becomes 1 count per frame in the COI dominated region.

\subsection{Coincidence loss correction}
\label{subsec:coincidence_loss_correction}
In order to correct for coincidence loss, it is necessary to designate an appropriate region for calculating the COI factor (COI region), because coincidence loss is an area effect and the shape of the COI region varies depending on source types.
For point sources, regardless of the radius of the photometric aperture (e.g., 5 or 3\,arcsec), a circular region with a radius of 5 arcsec, centered at the same position as the photometric aperture, is considered an appropriate COI region \citep{2008MNRAS.383..627P}.
Equations in \cite{2008MNRAS.383..627P} are applied to compute the COI factor for point sources:
\begin{equation}
    \begin{aligned}
        \dot{N}_{\rm theory}^{\rm COI}(\dot{N}_{\rm raw}^{\rm COI})&=\frac{-{\rm ln}(1-\alpha\dot{N}_{\rm raw}^{\rm COI}f_t)}{\alpha f_t} \\
        f(\dot{N}_{\rm raw}^{\rm COI})&=1+a_1 \dot{N}_{\rm raw}^{\rm COI} f_t+a_2(\dot{N}_{\rm raw}^{\rm COI}f_t)^2+a_3(\dot{N}_{\rm raw}^{\rm COI}f_t)^3+a_4(\dot{N}_{\rm raw}^{\rm COI}f_t)^4 \\
        {\rm COI}(\dot{N}_{\rm raw}^{\rm COI})&=f(\dot{N}_{\rm raw}^{\rm COI})\ \dot{N}_{\rm theory}^{\rm COI}(\dot{N}_{\rm raw}^{\rm COI})\ /\ \dot{N}_{\rm raw}^{\rm COI}
    \end{aligned}
    \label{equ:COI_point}
\end{equation}
where $\dot{N}_{\rm raw}^{\rm COI}$ is the raw/observed count rate of the point source COI region, $\alpha$ is the dead time correction factor 0.9842 (i.e. exposure time is 98.42$\%$ of full frame time), $f_t$ is the full frame time 0.0110329\,s.
$\dot{N}_{\rm theory}^{\rm COI}$ is the theoretical value of the COI corrected count rate in the point source COI region.
Function $f(x)$ is an empirical polynomial correction for the differences between theoretical and observed values.
Parameters $a_1$ to $a_4$ are 0.0669, -0.091, 0.029 and 0.031, respectively.
If the optimal photometric aperture of the UVOT is used to meeasure photometries , $\dot{N}_{\rm raw}^{\rm COI}$ equals the raw total count rate in the aperture $\dot{N}_{\rm raw}^{\rm tot}$, since the point source COI region is same as the optimal photometric aperture. While for extended sources, \cite{2015MNRAS.449.2514K} pointed out that rectangles with a typical area of $\sim400$ pixels are appropriate regions to calculate the COI factor with Equation (\ref{equ:COI_point}) for slitless spectra with different grism configurations obtained by UVOT.

However, neither of a circular or a rectangular region is suitable to calculate the COI factor for the wing.
Referring the method used by UVOT FTOOLS \citep{2014ApJ...790...52M} command {\it uvotsource} to calculate COI factors for backgrounds, which multiplies the raw background count rate density with the area of the point source COI region as the input value of Equation (\ref{equ:COI_point}) to calculate the COI factor, we want to find a region that has same/similar area as the point source COI region, where the raw count rate will be used for calculating the COI factor for the wing.
Therefore, the entire annulus is divided into 16 sector annuli, with each sector annulus having an opening angle of 22.5$^{\circ}$ and an area of $(25^2-15^2)\pi/16=25\pi$\,arcsec$^2$. Figure \ref{fig:COI_region} shows a kind of segmentation. Nevertheless, it is obvious the boundaries of these sector annuli can rotate around the center of the entire annulus with the assumption that counts distribute isotropically in the wing. Hence, it is recommended to use the mean value of raw total count rate of these sector annuli ($\dot{N}_{\rm raw,\,wing}^{\rm COI,\,tot}$) as the input value of Equation (\ref{equ:COI_point}) to calculate the COI factor for the wing:
\begin{equation}
    \dot{N}_{\rm raw,\,wing}^{\rm COI,\,tot}=\frac{25\pi[{\rm arcsec}^2]}{\rm area\ of\ unmasked\ wing}\times\dot{N}_{\rm raw,\,wing}^{\rm tot,\ UM}
    \label{equ:COI_wing}
\end{equation}
where $\dot{N}_{\rm raw,\,wing}^{\rm tot,\ UM}$ is the raw total count rate of unmasked wing region. Please pay much attention to the mask shape for the wing: Equation (\ref{equ:COI_wing}) only works correctly when masks are sector annuli that overlapped with sources in the wing, instead of shapes of sources (please refer to the last panel of Figure \ref{fig:example_measurement} for an example), because direct scaling measured count rate with area ratio is not correct for nonuniform sources.

However, Equation (\ref{equ:COI_point}) and (\ref{equ:COI_wing}) are only approximations for the true COI correction. Photons in the wing (especially inner edge) will be misrecognized with photons in the region surrounded by the wing (central photons later in short) when the wing is bright enough, but the COI correction employed in this article can not handle the case. In another words, when the region heavily influenced by the COI effect from central photons expands to the wing, the PSF method can not be applied to restore photometries any more.

\subsection{Additional correction for extended sources}
The method employed by UVOT FTOOLS command {\it uvotsource} can not completely correct the coincidence loss for backgrounds \citep{2010MNRAS.406.1687B}.
In other words, to obtain the true incident count rate for uniform extended sources, the raw count rate needs to be multiplied with an additional correction factor, which is named as the Extended (EXT) correction factor.
The EXT correction essentially serves as an additional correction for the coincidence loss, due to the inadequate COI factor for extended sources.
Therefore, the equivalent raw count rate, scaled with the area ratio of the point source COI region to the measuring region for uniform extended sources ($\dot{N}_{\rm raw,\,ext}^{\rm COI}$), is also used to calculate the EXT factor, which is fitted phenomenologically using data points from Figure 6 of \cite{2010MNRAS.406.1687B} and a smoothed broken power-law model:
\begin{equation}
    {\rm EXT}=(1+(\dot{N}_{\rm raw,\,ext}^{\rm COI}/\dot{N}_{\rm 0,\,ext}^{\rm COI})^{a})^b
    \label{equ:EXT}
\end{equation}
Best fitted parameters are $\dot{N}_{\rm 0,\,ext}^{\rm COI}=160.115922$\,count/s, $a=1.518061$ and $b=2.446816$. Figure \ref{fig:EXT_factor} shows the difference between true incident and COI corrected count rate, and the fitting result for the EXT factor.

For typical background values of UVOT \citep[$0-0.05$\,count/s/pixel,][]{2010MNRAS.406.1687B}, the background is brightened by $\sim7\%$ at most.
Its effect on photometries for point sources is so small that can be neglected safely, because the EXT factor is only multiplied with the raw background count rate, which is usually much smaller than the count rate of the point source.
While for extended sources, the EXT correction becomes important since both the wing and the background need to be corrected. The non-uniformity is worthy being taken into account, but actually, it only has little influence on the PSF method and can be neglected safely. Please refer to Appendix \ref{app:NU} for the detailed discussion.

\subsection{Large scale structure and sensitivity correction}
Large Scale Structure (LSS) and SENsitivity (SEN) correction should be applied to COI corrected count rates to get the final corrected count rate that can be used for the photometry. The LSS depends on the source position on the detector and the bandpass, and the sensitivity correction (SEN) only depends on the time when the observation was executed. LSS and SEN can be found with UVOT FTOOLS, e.g. photometry table generated by {\it uvotsource} command. Hence, the corrected source count rate in the wing ($\dot{N}_{\rm wing}$) can be derived with:
\begin{equation}
    \begin{aligned}
        \dot{N}_{\rm CE,\,wing}^{\rm tot}&=\frac{\rm area\ of\ entire\ wing}{\rm area\ of\ unmasked\ wing}\times\dot{N}_{\rm raw,\,wing}^{\rm tot,\,UM}\,{\rm COI}_{\rm wing}^{\rm tot}\,{\rm EXT}_{\rm wing}^{\rm tot} \\
        \dot{N}_{\rm CE,\,wing}^{\rm bkg}&=400\pi\,[{\rm arcsec}^2]\times{\rm CRD}_{\rm raw}^{\rm bkg}\,{\rm COI}_{\rm wing}^{\rm bkg}\,{\rm EXT}_{\rm wing}^{\rm bkg} \\
        \dot{N}_{\rm wing}&=(\dot{N}_{\rm CE,\,wing}^{\rm tot}\,-\,\dot{N}_{\rm CE,\,wing}^{\rm bkg})\times{\rm LSS}\times{\rm SEN}
    \end{aligned}
    \label{equ:N_wing}
\end{equation}
where $\dot{N}_{\rm CE,\,wing}^{\rm tot}$ and $\dot{N}_{\rm CE,\,wing}^{\rm bkg}$ are the COI and EXT corrected total and background count rates in the wing. At this stage, it can be concluded that the inadequate correction for the COI (when the COI effect from central photons is important), and the intrinsic non-uniformity of the wing contribute to the theoretical systematic uncertainty of the PSF method. However, it is hard to quantify the exact value of them, hence the systematic uncertainty of the PSF method is estimated with the fluctuation of calibration residuals, i.e., the comprehensive systematic uncertainty involving both theoretical (measuring principle) and practical (calibration data set) ones. Please refer to Section \ref{subsubsec:calibration_of_zp} and Table \ref{tab:ZP}.

\section{Saturated correction method}
\subsection{The principle of the calibration}
If the profile of the PSF is known, the point source count $C_{\rm src}$ can be derived by fitting Growth Curve (GC) of the PSF: $C_{\rm src}GC(r)=C_{\rm src}\int_{0}^{2\pi}\int_{0}^{r}PSF(r',\theta)\times r'd\theta dr'$, where $PSF(r,\theta)$ is the normalized profile of the PSF (i.e., $\int_{0}^{2\pi}\int_{0}^{\infty}PSF(r,\theta)\times rd\theta dr=1$), and $C_{\rm src}$ is the source count.
Usually, integrating to a standard or optimal radius instead of infinity is more practical, since there would be many sources in large integrating area that will interfere the integration. In addition, if the PSF is stable, the ratio of count rate in standard aperture to true count rate keeps same, i.e. $\int_{\rm std}PSF(r,\theta)\times rd\theta dr/\int_{\rm inf}PSF(r,\theta)\times rd\theta dr=constant$.
As a result, flux calibration can base on count rate in standard aperture. For UVOT, the radius of standard photometric aperture suggested in \cite{2008MNRAS.383..627P} is 5\,arcsec. For saturated stars, the PSF profile is destroyed near $r=0$, but the wing of the PSF becomes bright enough to make relatively accurate measurement, and the count rate in the wing can be converted to count rate in the standard region by simply multiplying a factor $k=\int_{\rm std}PSF(r,\theta)\times rd\theta dr/\int_{\rm wing}PSF(r,\theta)\times rd\theta dr$.

Figure.5 in \cite{2008MNRAS.383..627P} shows the PSF profile of UVOT, assuming it is isotropic. The wing of the PSF extends to a radius of $\sim$ 50 pixels, and as shown in Figure \ref{fig:saturation_pattern}, the COI strongly influences the area within $\sim15$\,arcsec radius. Hence, the wing of UVOT's PSF is defined as an annulus with an inner radius of 30 pixels (15\,arcsec) and an outer radius of 50 pixels (25\,arcsec).
 If the PSF of UVOT is stable, $\dot{N}_{\rm std}$ should be proportional to $\dot{N}_{\rm wing}$, i.e. $\dot{N}_{\rm std}=k\dot{N}_{\rm wing}$, where $\dot{N}_{\rm wing}$ and $\dot{N}_{\rm std}$ represent the count rate in wing and the count rate in the standard photometric region, respectively.
The $\dot{N}_{\rm std}$ can be converted to the AB magnitude with the equation $M^{\rm AB}=-2.5{\rm lg}(\dot{N}_{\rm std}[{\rm count/s}])+ZP^{\rm AB}_{\rm std}$, where the $ZP^{\rm AB}_{\rm std}$ is the photometric zeropoint in AB system if a standard photometric aperture is used. For UVOT, $ZP^{\rm AB}_{\rm U, std}=19.36\pm0.02$, $ZP^{\rm AB}_{\rm B, std}=18.98\pm0.02$, $ZP^{\rm AB}_{\rm V, std}=17.88\pm0.01$ \citep{2008MNRAS.383..627P,2011AIPC.1358..373B}. Hence the AB magnitude of a source can be calculated with $\dot{N}_{\rm wing}$ by equation:
\begin{equation}
    \begin{aligned}
        M^{\rm AB}&=-2.5{\rm lg}(k\dot{N}_{\rm wing}[{\rm count/s}])+ZP^{\rm AB}_{\rm std} \\
        &=-2.5{\rm lg}(\dot{N}_{\rm wing}[{\rm count/s}])+ZP^{\rm AB}_{\rm wing}
    \end{aligned}
    \label{equ:AB_wing}
\end{equation}
where $ZP^{\rm AB}_{\rm wing}=ZP^{\rm AB}_{\rm std}-2.5{\rm lg}(k)$. $ZP_{\rm wing}^{\rm AB}$ is the only free parameter of Equation (\ref{equ:AB_wing}) needed to be calibrated.

\subsection{AB magnitudes of saturated {\rm UVOT} sources converted from other catalogues}
The crucial problem is how to get reliable $M^{\rm AB}$ of saturated UVOT sources. One solution is using the color transformation between the UVOT filter system and other filter systems to get converted $M^{\rm AB}$ for saturated UVOT sources.

\subsubsection{Gaia Synthetic Photometry Catalogue}
The Gaia Synthetic Photometry Catalogue (GSPC) provides synthetic photometris for some widely used filter systems \citep[e.g., the Johnson-Kron-Cousin system defined by][]{2012PASP..124..140B}, which are generated from spectra with high signal-to-noise ratio($>$ 30) in Gaia Data Release 3, and contains $\sim$ 220 million sources \citep{2022arXiv220606215G}.
As shown in Figure \ref{fig:VBU_Comparison}, the effective transmission curve of UOVT VBU bands (${\rm VBU}_{\rm UVOT}$) are very similar to that of Johnson-Kron-Cousin VBU bands (JKC in following article, ${\rm VBU}_{\rm JKC}$), hence the synthetic photometries of VBU$_{\rm JKC}$ are selected to be converted to VBU$_{\rm UVOT}$.
Effective transmission curves of JKC system are downloaded from the ftp of CDS\footnote{\url{https://cdsarc.cds.unistra.fr/viz-bin/cat/J/PASP/124/140}}, and effective transmission curves of UVOT system are taken from the \textit{Swift} CALDB website\footnote{\url{https://heasarc.gsfc.nasa.gov/docs/heasarc/caldb/swift/docs/uvot/uvot_caldb_filtertransmission_03.pdf}}.
Since photometries of Johnson-Kron-Cousin system are usually reported in Vega system, transformation between VBU$_{\rm JKC}$ and VBU$_{\rm UVOT}$ is derived under Vega system. The Vega magnitude of a star is derived with the following equation:
\begin{equation}
M^{\rm Vega}=\frac{\int T(\lambda)\lambda F_{\lambda}(\lambda)d\lambda}{\int T(\lambda)\lambda F_{\lambda}^{\rm Vega}(\lambda)d\lambda}+ZP^{\rm Vega},
\end{equation}
where $\lambda$ is wavelength, $T(\lambda)$ is the effective transmission curve, $F_\lambda(\lambda)$ and $F_\lambda^{\rm Vega}(\lambda)$ are flux density of observed source and flux density of Vega in unit of flux per wavelength, respectively. $M^{\rm Vega}$ represents this magnitude is expressed in the Vega system and $ZP^{\rm Vega}$ is the zeropoints (i.e., the magnitude of Vega in this system). For UVOT, $ZP^{\rm Vega}_{\rm UVOT}=0$\footnote{\url{https://heasarc.gsfc.nasa.gov/docs/heasarc/caldb/swift/docs/uvot/uvot_caldb_zeropoints_10wa.pdf}}, however, in GSPC, $ZP^{\rm Vega}_{\rm GSPC}=0.03$ \citep{2014AJ....147..127B,2022arXiv220606215G}, since the reference spectrum $F_\lambda^{\rm Vega}(\lambda)$ is not completely same as the most accurate observed spectrum.
We measured synthetic photometrires of stars in \cite{1998PASP..110..863P} stellar library, except M type stars, because the scatter of M type stars is much large than other stars, which is also found by \cite{2013MNRAS.436.1684P} and \cite{1997yCat.1239....0E}. Since there are 3 colors, i.e. U$_{\rm JKC}$ - B$_{\rm JKC}$, U$_{\rm JKC}$ - V$_{\rm JKC}$ and B$_{\rm JKC}$ - V$_{\rm JKC}$, could be converted to VBU$_{\rm UVOT}$ - VBU$_{\rm JKC}$, every possible transformation was fitted with a piecewise linear function by least square method. 
However, only the fit with the smallest residual is selected as the final color transformation. For example, there are 3 possible transformations for U$_{\rm UVOT}$ - U$_{\rm JKC}$, i.e. U$_{\rm UVOT}$ - U$_{\rm JKC}$ versus U$_{\rm JKC}$ - B$_{\rm JKC}$, U$_{\rm JKC}$ - V$_{\rm JKC}$ or B$_{\rm JKC}$ - V$_{\rm JKC}$, and Root Mean Squares (RMSs) are 0.0316, 0.0327 and 0.03861, respectively. The RMS of U$_{\rm UVOT}$ - U$_{\rm JKC}$ versus U$_{\rm JKC}$ - B$_{\rm JKC}$ is the smallest, hence it is selected as the final color transformation for U band. The color transformation is concluded in following equations and shown in Figure \ref{fig:color_transformation}.
\begin{equation}
\begin{aligned}
    {\rm U_{UVOT} - U_{JKC}}&=
    \begin{cases}
        {\rm 0.189(U_{JKC} - B_{JKC}) - 0.054},\ {\rm U_{JKC} - B_{JKC}} \leq 0.079 \\
        {\rm 0.018(U_{JKC} - B_{JKC}) - 0.040},\ {\rm U_{JKC} - B_{JKC}} > 0.079
    \end{cases} \\
    {\rm B_{UVOT} - B_{JKC}}&=
    \begin{cases}
        {\rm 0.007(B_{JKC} - V_{JKC}) - 0.031},\ {\rm B_{JKC} - V_{JKC}} \leq 1.192 \\
        {\rm 0.085(B_{JKC} - V_{JKC}) - 0.123},\ {\rm B_{JKC} - V_{JKC}} > 1.192 
    \end{cases} \\
    {\rm V_{UVOT} - V_{JKC}}&=
    \begin{cases}
    {\rm 0.011(B_{JKC} - V_{JKC}) - 0.036},\ {\rm B_{JKC} - V_{JKC}} \leq 0.167 \\
    {\rm 0.038(B_{JKC} - V_{JKC}) - 0.041},\ {\rm B_{JKC} - V_{JKC}} > 0.167
    \end{cases}
\end{aligned}
\end{equation}
RMSs for V, B, and U bands are 0.0032, 0.0056, and 0.0316, respectively, which are treated as the systematic uncertainties of color transformation. The color transformation is in Vega system.
With the definition of AB and Vega magnitude, the AB magnitude of a source equals the Vega magnitude of it plus the AB magnitude of Vega in the same band, i.e. $M^{\rm AB}_{\rm Band}=M^{\rm Vega}_{\rm Band}+M^{\rm AB}_{\rm Vega,Band}$.
According to \cite{2011AIPC.1358..373B}, $M^{\rm AB}_{\rm Vega,V_{UVOT}}=-0.01$, $M^{\rm AB}_{\rm Vega,B_{UVOT}}=-0.13$, $M^{\rm AB}_{\rm Vega,U_{UVOT}}=1.02$.

\subsubsection{Tycho-2 catalogue}
In the Tycho-2 catalogue, there are two-color photometric data of 2.5 million brightest stars \citep{2000A&A...355L..27H}. \cite{2013MNRAS.436.1684P} studied the color transformation between Tycho-2 and UVOT, and used Tycho-2 sources to calibrate the readout streak method to increase the measurement range of UVOT for V and B bands. Hence, Tycho-2 sources are also used to calibrate the PSF method. The color transformation between Tycho-2 and UVOT is taken from \cite{2013MNRAS.436.1684P}:
\begin{equation}
    \begin{aligned}
        {\rm V}_{\rm UVOT}&={\rm V}_{\rm T}-0.032-0.073({\rm B}_{\rm T}-{\rm V}_{\rm T}),\,{\rm B}_{\rm T}-{\rm V}_{\rm T} > 0 \\
        {\rm B}_{\rm UVOT}&={\rm B}_{\rm T}+0.036-0.270({\rm B}_{\rm T}-{\rm V}_{\rm T}),\,{\rm B}_{\rm T}-{\rm V}_{\rm T} > 0.4
    \end{aligned} 
\end{equation}
where V$_{\rm T}$ and B$_{\rm T}$ are magnitudes in Tycho-2 system. Note that this transformation is also in Vega system. RMSs are 0.006 and 0.009 and for V and B bands, respectively.

\cite{2005PASP..117..615M} pointed out that magnitudes recorded in Tycho-2 catalogue are slightly brighter than true values, for V$_{\rm T}$ is $0.058\pm0.009$ mag and for B$_{\rm T}$ is $0.078\pm0.009$ mag, which also is found in our sample: UVOT magnitudes converted from Tycho-2 sources are slightly brighter than that converted from GSPC sources, $<{\rm V}^{\rm Tycho2}_{\rm UVOT}-{\rm V}^{\rm GSPC}_{\rm UVOT}>=-0.0485$ and $<{\rm B}^{\rm Tycho2}_{\rm UVOT}-{\rm B}^{\rm GSPC}_{\rm UVOT}>=-0.0576$, where $M^{\rm CAT}_{\rm UVOT}$ represents UVOT magnitudes converted from sources of ``CAT" catalogue. The bias can be significantly reduced to 0.008 and 0.015 and for V and B bands if Tycho-2 magnitudes were corrected with values from \cite{2005PASP..117..615M} before the color transformation.

\subsection{Generation of calibration data}
\label{subsec:generation_of_calibration_data}
\subsubsection{Process of UVOT data}
\label{subsubsec:process_of_uvot_data}
First, the UVOT observation catalogue {\it swiftuvlog} (up to March, 2023) is searched with parameters: obs\_segment=000, operation\_mode=EVENT and asp\_corr=Y.
For the first criterion, obs\_segment means the number of times a specific target has been exposed (e.g., usually, 000 represents the first exposure sequence).
For the second criterion, we hope our sample was operated under event mode, and the data set will not be too large.
In addition, UVOT usually works under event mode at the early stage of burst events like Gamma-Ray Bursts (typically several hundred seconds since the trigger), so there is image mode data in the selected sample as well.
For the last one, the asp\_corr represents whether the aspect correction was applied to the data, in other words, it can be simply understood as whether the sky image is well aligned with the WCS. 
There are total 1851, 1411 and 1373 observations were selected for V, B and U bands, respectively.

Next, sky and exposure images for each observation were downloaded and a Pre-Process (PP) algorithm was applied to them.
There are 3 main purpose for the pre-process algorithm: 1) drop bad extensions, e.g. duplicate extension names and nonuniform exposure/elapsed time. 2) classify extensions by their operation modes and bin factor into 4 groups, i.e. event 1x1, event 2x2, image 1x1 and image 2x2. 3) stack all good extensions for a deep exposure of the observation (called as stacked PP image in following article).

\subsubsection{Process of Tycho-2 catalogue and GSPC data}
\label{subsubsec:process_of_tycho_2_catalogue_and_gspc_data}
Tycho-2 and GSPC catalogue are cross matched with selected UVOT observation list with a search radius of 12 arcmin. For Tycho-2 catalogue, there are 12041 and 6873 sources in V and B bands. For GSPC, there are 2094, 2002 and 1954 sources for V, B and U bands. Then, all sources are filtered with each extension of corresponding PP images by 3 criteria: 
\begin{itemize}
    \item[1)] Does the color of the source fall in the valid range for color transformation? 
    \begin{itemize}
        \item[$\bullet$] $-0.5<{\rm B}_{\rm JKC} - {\rm V}_{\rm JKC}<2$ for GSPC V-band sources
        \item[$\bullet$] $-0.5<{\rm B}_{\rm JKC} - {\rm V}_{\rm JKC}<2$ for GSPC B-band sources
        \item[$\bullet$] $-1.5<{\rm U}_{\rm JKC} - {\rm B}_{\rm JKC}<2$ for GSPC U-band sources
        \item[$\bullet$] $0<{\rm B}_{\rm T} - {\rm V}_{\rm T}<2$ for Tycho-2 V-band sources
        \item[$\bullet$] $0.4<{\rm B}_{\rm T} - {\rm V}_{\rm T}<2$ for Tycho-2 B-band sources
    \end{itemize}
    \item[2)] Is the wing of the source intact in the field of UVOT?
    \item[3)] Is the wing of the source exposed uniformly?
\end{itemize}
Meanwhile, for each source that passes the 3 criteria, 4 files are generated: photometry information for the source (including the converted $M^{\rm AB}$ for saturated UVOT sources and will be used to generate the catalogue of filtered good stars), ds9 region file, filtered images and a ``stack\&phot" script for stacking and photometry. Finally, for Tycho-2 catalogue, there are [1821, 166, 1079] and [1039, 27, 591] stars left for [event 1x1, image 1x1, image 2x2] modes in V and B bands. For GSPC, there are [419, 33, 249], [326, 2, 176] and [357, 251, 177] stars left for [event 1x1, image 1x1, image 2x2] modes in V, B and U bands. The last step is to measure $\dot{N}_{\rm wing}$ for each saturated source. Figure \ref{fig:pipeline} shows the flow chart for all steps mentioned in Section \ref{subsubsec:process_of_uvot_data} and \ref{subsubsec:process_of_tycho_2_catalogue_and_gspc_data}.

\subsubsection{Measuring the source count rate of the wing}
\label{subsubsec:measurement_of_N_wing}
The ``stack\&phot" script will run UVOT FTOOLS to stack filtered images and make photometry with the standard photometric aperture of UVOT. LSS and SEN for each source are read from the fits table of photometry results. With the stacked PP image, sources in the field of the stacked filtered image are marked, so the property of the background could be estimated.

Since the wing extends up to 25\,arcsec, there could be other sources in the wing. How to mask possible sources in the wing is an important problem, because normal source detection algorithm can not detect point sources in the wing efficiently. Hence, the pattern of the wing should be subtract before the source detection. Images of about 30 isolated stars are stacked to derive the template of the Count Density Distribution (CDD, counts per unit area at a specific radius) of the wing for V, B and U bands.
The 0th, 1st and 2rd orders of Chebyshev polynomials are used to fit the template of CDD, for a reference, [0th, 1st, 2rd coefficients] are [2.193856, -0.070483, -0.000234], [0.965346, 0.064040, -0.001984] and [2.756965, -0.128014, 0.000472] for V, B and U bands, respectively. The CDD template is only a proper approximation for cases $\dot{N}_{\rm raw,\,wing}^{\rm tot}<\sim60$\,count/s. For cases $\dot{N}_{\rm raw,\,wing}^{\rm tot}>\sim60$, the wing pattern would be over subtracted, because the CDD depends on the $\dot{N}_{\rm raw,\,wing}^{\rm tot}$ due to the coincidence loss. Therefore, fitted CDDs of saturated sources can not be used to calculate $\dot{N}_{\rm wing}$ directly. Figure \ref{fig:COI_region} shows the V-band stacked images and Figure \ref{fig:CDD} shows fitted templates of CDDs.

Once the pattern of the wing is subtracted, the sigma clip method is applied to the residual image to identify possible sources and generate a bad pixel map. Pixels rejected by the sigma clip method is marked as 1 and others are 0 in the bad pixel map. Then the bad pixel map is smoothed with a uniform kernel ($3\times3$, and each element is 1/9), and only pixel with value greater than 0.5, (i.e., at least 5 pixels are rejected in regions of $3\times3$ pixels), are marked as true sources to avoid rejection of true extreme background values.
Finally, a sector annulus with an opening angle of 10$^{\circ}$ and same inner and outer radius of the wing is used to scan the entire wing region. If there is any rejected pixel in the sector annulus, the sector annulus will be marked as mask region, instead of the shape of the source itself. With Equation (\ref{equ:COI_point}-\ref{equ:EXT}, and \ref{equ:N_wing}) and the mask map, $\dot{N}_{\rm wing}$ finally can be measured. An example for the measurement is shown Figure \ref{fig:example_measurement}.

\subsection{Calibrating photometric zeropoints of the PSF method}
\label{subsubsec:calibration_of_zp}
Before further analysis, all data with signal-to-noise ratio less than 10 is dropped. For GSPC data, $M^{\rm AB,\ GSPC}_{\rm UOVT}$ has obvious lower and upper boundaries, but these does not affect the result of $ZP^{\rm AB}_{\rm wing}$ a lot, because both lower and upper boundaries are complete. While for Tycho-2 data, $M^{\rm AB,\,Tycho2}_{\rm UOVT}$ has a valid but not obvious lower boundary, and $\dot{N}_{\rm wing}$ reaches $\sim700$ count/s, which is far beyond the range the PSF method can be applied.
Hence, for V band, only Tycho-2 sources with $15\,{\rm count/s}<\dot{N}_{\rm wing}<100\,{\rm count/s}$ are fitted and for B band is $25\,{\rm count/s}<\dot{N}_{\rm wing}<100\,{\rm count/s}$. In addition, because uncertainties of GSPC data ($\sim0.005$ mag) are much smaller than that of Tycho-2 data ($\sim0.2$ mag for low $\dot{N}_{\rm wing}$ and $\sim0.05$ for high $\dot{N}_{\rm wing}$) in V and B bands, there are additional scaling factors for GSPC data when calculating their cost to ensure that Tycho-2 and GSPC sources have similar effects on the fitting, see Figure \ref{fig:V_res_dis}. To summary, all filtered GSPC data and a part of filtered Tycho-2 data were fitted with Equation (\ref{equ:AB_wing}).

Another import thing is that there are outliers for the fitting, especially those points significantly lower than the most data. To pick up outliers, we first fitted all points with the least square method and used the criterion to select candidates of outliers: residuals greater than 3 times RMS of residuals or quantiles of cost values less than 10\% (for points brighter than the fitted model) and greater than 80\% (for points fainter than the fitted model) of all cost values. 
Images of all candidates were checked and we found there is no physical reason to mark most bright candidates as outliers (only a few, 7 of 111 checked sources for Tycho-2 V-band event 1x1 data), while most faint candidates should be marked as outliers ($\sim100$ of 156 checked sources for Tycho-2 V-band event 1x1 data). Faint outliers are mainly marked for 2 reasons: 1) the source is actually unresolved binary stars by UVOT (but resolved in Tycho-2 catalogue or GSPC). 2) the wing of the source is contaminated by halo ring, diffraction spikes or other extended sources \citep{2014styd.confE..37P}.
Hence, it is confident to mark these faint points as outliers. In addition, if the source is in a very crowed field or its sky coordinate is not accurate, it is marked as an outlier as well (e.g., most bright outliers). Total [1241, 111, 673], [648, 11, 348] and [250, 213, 145] stars are finally fitted for [event 1x1, image 1x1, image 2x2] modes in V, B and U bands. Fitting results without the consideration of different operation modes and bin factors for images are shown in Figure \ref{fig:V_all}-\ref{fig:U_all}. Calibrated $ZP^{\rm AB}_{\rm wing}$ are summarized in Table \ref{tab:ZP}. Operation modes and bin factors for images do not influence the photometries obtained by the PSF method. For calibration of different operation modes and image bin factors, please refer to Appendix \ref{app:Individual_Calibration}.

The PSF method only can be applied in valid ranges for $\dot{N}_{\rm wing}$. As mentioned above, for wings brighter than upper boundary of the valid range, the COI effect from central photons on the wing becomes important and can not be neglected for V and B bands. Figure \ref{fig:V_all} and Figure \ref{fig:B_all} show that when $\dot{N}_{\rm wing}>\sim120~$count/s, the calibration data points deviate from the model, and the reason is Equation \ref{equ:COI_point} and \ref{equ:COI_wing} can not correct the COI effect from central photons properly, as a result, their $\dot{N}_{\rm wing}$ are smaller than true value. The evidence can be identified from the image. As the panel (b) in Figure \ref{fig:saturation_pattern} shows, the wing is influenced by the COI effect from central photons (the small white spot at the lower left part of the wing is exactly contaminated by the COI pattern from central photons). While the upper boundary for U band is limited by the brightest calibration sources. The lower boundary is set by the faintest calibration sources for all bands. For V band, $10\,{\rm count/s}<\dot{N}_{\rm wing}<100\,{\rm count/s}$. For B band, $20\,{\rm count/s}<\dot{N}_{\rm wing}<100\,{\rm count/s}$. The a bit higher lower threshold (15\,count/s and 25\,count/s for V and B bands) is set to avoid the influence of the lower boundary of the fitted data set, hence the lower limit of valid range could be slightly lower. For U band, $12\,{\rm count/s}<\dot{N}_{\rm wing}<40\,{\rm count/s}$. Residuals of fitted data are divided into several bins according to their $\dot{N}_{\rm wing}$, and quantile values of 68.26\%, 95.44\% and 99.74\% of absolute residuals are regarded as 1-$\sigma$, 2-$\sigma$ and 3-$\sigma$ uncertainties of $ZP^{\rm AB}_{\rm wing}$ for each bin. The mean value of 1-$\sigma$ uncertainties of $ZP^{\rm AB}_{\rm wing}$ for each bin (except bins have less than 25 data points or influenced by the lower/upper boundary of fitted data set) is regarded as the 1-$\sigma$ systematic uncertainty of $ZP^{\rm AB}_{\rm wing}$ for the entire valid range of $\dot{N}_{\rm wing}$ ($\sim0.2$ mag for V, B and U bands, please see Table \ref{tab:ZP} for exact numbers).

However, only $\sim30\%$ images of all fitted data are checked carefully, and there are still unmarked outliers in the fitted data set. In addition, most outliers are lower than the fitted model, hence the fitted $ZP^{\rm AB}_{\rm wing}$ is expected to be slightly fainter. Based on the fact that all fitted $ZP^{\rm AB}_{\rm wing}$ became $\sim0.1$ mag brighter after removing outliers (especially some have extreme cost values) from fitted data, an estimation for the bias is $<\sim0.1$ mag.

There is no special restriction for the PSF method regarding the position of the saturated source. For a reference, the spatial distribution of Tycho-2 sources used to calibrate $ZP_{\rm V,\,wing}^{\rm AB,\,evt1x1}$ is plotted in Figure \ref{fig:cal_src_dis_V_evt1x1}.  However, just keep sure the wing is exposed uniformly. We found there is a gradient trend for the elapsed time in the field of view by checking the exposure image in some observations (rare but indeed exist).

\section{Demonstration}
\label{sec:Demonstration}
The PSF method is applied to restore photometries of the famous ``naked eye" GRB 080319B early saturated UVOT observations in V band. The ground-based telescope RAPTOR-T \citep{2009ApJ...691..495W} well observed GRB 080319B at early phase in V band, when it was being observed by UVOT under event mode in V band.
The event file is screened into sub exposures that match with the exposure sequence of RAPTOR-T.
Following steps described in Section \ref{subsubsec:measurement_of_N_wing}, count rates of the wing $\dot{N}_{\rm wing}$ are measured for each screened sub exposure and corresponding magnitudes are calculated with Equation (\ref{equ:AB_wing}) with the V-band zeropoint of the PSF method, see Table \ref{tab:ZP}.
The PSF method is also applied to the 4 time bins screened by \cite{2013MNRAS.436.1684P}.
Table \ref{tab:080319B_V} lists measurements of the early V-band observations of GRB 080319B and Figure \ref{fig:GRB_lc} shows the comparison between measurements of RAPTOR-T, UVOT pipeline, the readout streak method, and the PSF method.
It is clearly to be seen, before 400\,s after the trigger, most RAPTOR-T points are higher than PSF method points by $\sim0.1$ mag, while after 400 s, RAPTOR-T points are well matched with PSF method points and UVOT pipeline measurements.
Here are 3 possible reasons to explain this: 1) residual outliers in the fitted data set, 2) the intrinsic difference between RAPTOR-T V and UVOT V, 3) UVOT measurements could be slightly fainter than true values when the source is barely saturated in V band. For the detailed explanation, please refer to Appendix \ref{app:Difference_between_UVOT_and_JKC}.
Anyway, measurements of RAPTOR-T are consistent with that of the PSF method within 1-$\sigma$ uncertainty and indeed the $ZP^{\rm AB}$ derived with our sample is slightly underestimated at a level of $<\sim0.1$ mag, but it can be solved by using better saturated sources (e.g., isolated and have spectra) for the calibration.
Photometries restored by the readout streak method are slightly brighter than $\sim0.1$ mag, but also consistent with RAPTOR-T points within 1-$\sigma$ uncertainty.
It can be found uncertainties of the PSF method are smaller than that of the readout streak method for barely saturated sources, hence the PSF method can be applied to images with shorter exposure time. However, the readout streak method can restore strongly saturated sources, to which the PSF method can not be applied.

GRB 210702A has a bright optical counterpart and 315 seconds after the BAT trigger, the UVOT took a 45\,s exposure in U band. The preliminary U-band photometry result reported by \textit{Swift}/UVOT team is 11.37$\pm$0.05 mag in Vega system \citep{2021GCN.30356....1K}. However, the saturation magnitude of U-band is 11.91th mag (Vega), hence the optical transient actually saturated in the 45s U-band image. We used the source to test the PSF method. The $\dot{N}_{\rm wing}$ is 33.75$\pm$1.42\,count/s and the corrected saturated photometry is $12.34\pm0.05\pm0.17$th mag (AB), which corresponds to $11.32\pm0.18$th mag (Vega) and consistent with the value $11.37\pm0.05$ given by \textit{Swift}/UVOT team. When measuring $\dot{N}_{\rm wing}$, there is a faint smoke ring \citep{2014styd.confE..37P} at the bottom right corner of the annulus region, and a sector annulus region with an opening angle of 90$^{\circ}$ was used to mask it.

\section{Conclusion}
\begin{figure}[ht]
    \centering
    \includegraphics[width=0.8\columnwidth]{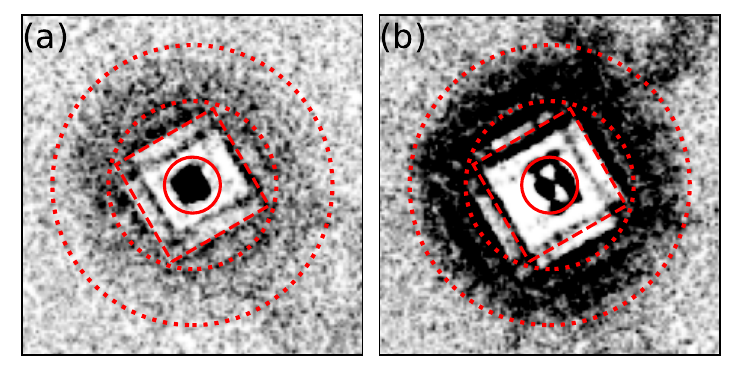}
    \caption{Saturation pattern of UVOT images and definitions for the core and the wing regions. In both panels, the solid red circle with a radius of 5\,arcsec (the core of the PSF) is the standard/optimal photometric aperture of UVOT for point sources and also the point source COI region. The dotted annulus with an inner radius of 15\,arcsec and an outer radius of 25\,arcsec is defined as the wing of the PSF. The dashed red square represents with a side length of 20\,arcsec represents the region heavily influenced by the coincidence loss (white square pattern) for moderate saturation. For panel (a), the raw total count rate of the wing $\dot{N}_{\rm raw,\,wing}^{\rm tot}$ is $\sim60\,$count/s, corresponding to the case of moderate saturation, and the COI dominated region is in the dashed square. While panel (b) shows a strongly saturated case, where $\dot{N}_{\rm raw,\,wing}^{\rm tot}$ is $\sim150\,$count/s, and the COI dominated region expands beyond the dashed square region. The extended source at the upper right corner of panel (b) is an artifact \citep[smoke ring,][]{2014styd.confE..37P} caused by nearby real bright point sources. The color scale for both panels are same.} 
    \label{fig:saturation_pattern}
\end{figure}

\begin{figure}[h]
    \centering
    \includegraphics[width=0.4\columnwidth]{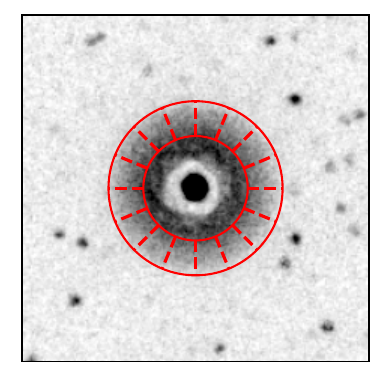}
    \caption{The stacked image of isolated stars in V band and one kind of segmentation of COI regions for the wing. The solid red annulus is the wing defined before. Dashed red lines are boundaries of each sector annulus with an opening angle of 22.5$^\circ$. The boundaries can rotate around the center of the entire annulus with the assumption that counts distribute isotropically in the wing, hence how to set them is arbitrary.} 
    \label{fig:COI_region}
\end{figure}

\begin{figure}[ht]
    \centering
    \includegraphics[width=0.4\columnwidth]{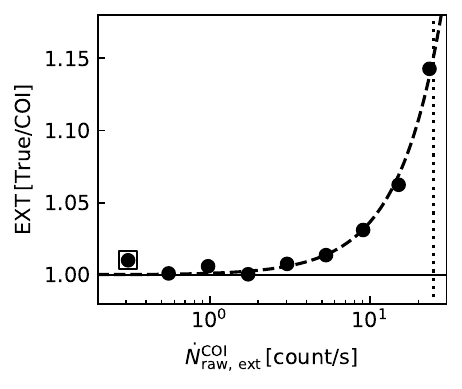}
    \caption{The additional adjustment of coincidence loss for extended sources. The horizontal axis represents the scaled raw count rate with the area ratio of the point source COI region (25$\pi\,$[arcsec$^2$]) to the measuring region for uniform extended sources. The vertical axis represents the EXT factor (the quotient of the true incident count rate divided by the COI corrected count rate). Data is taken from Figure 6 of \cite{2010MNRAS.406.1687B}. The point marked with a square is excluded for the fitting procedure due to its abnormal deviation from the expected value. Both the COI and EXT factors are expected to be 1 when the effect of the coincidence loss is negligible (i.e., $\dot{N}_{\rm raw,\,ext}^{\rm COI}$ is $\sim0\,$count/s), but compared with other points, the excluded point deviates unreasonably from the expected value, which might be caused by the undisclosed large uncertainty. The vertical dotted line represents the valid threshold for the fitted model, i.e. $\dot{N}_{\rm raw,\,ext}^{\rm COI}<25\,$count/s.} 
    \label{fig:EXT_factor}
\end{figure}

Although the imaging principle of UVOT is different from usual optical telescopes with a CCD, e.g. HST, it is still able to derive photon counts of saturated sources with photon counts in wing regions of the PSF. However, as shown in Figure \ref{fig:saturation_pattern}, the region dominated by the coincidence loss is very large, hence the wing of the PSF defined in this paper is relatively large (an annulus ranging from 15\,arcsec to 25\,arcsec) to avoid the influence from the coincidence loss of the central core, which makes the measurement of $\dot{N}_{\rm wing}$ challenging, especially there are sources in the wing. However, by subtracting the wing pattern with derived CDD template, sources in the wing region can be marked efficiently with smoothed the bad pixel map generated with the sigma clip method, which makes measuring $\dot{N}_{\rm wing}$ of a relatively large sample possible.
Tycho-2 and GSPC sources are used to calibrate photometric zeropoints of the PSF method. Just to be sure, the calibration data set for each band is divide into 3 sub sets by different operation modes and image bin factors for calibration, to check if they have influence on photometric zeropoints of the PSF method. As shown in Table \ref{tab:ZP}, UVOT operation modes and image bin factors have no influence on photometric zeropoints of the PSF method, hence just using the zeropoint for a specific band is reasonable.

For a reference, the steps to use the PSF method is summaried:
\begin{itemize}
    \item[1)] Keep sure the wing is exposed uniformly by checking the exposure map.
    \item[2)] If there any source or artifact in the wing, use serveral small sector annuli overlapped with all sources as the mask, see last panel in Figure \ref{fig:example_measurement} for a reference.
    \item[3)] Measure the raw total count rate of the unmasked wing ($\dot{N}_{\rm raw,\,wing}^{\rm tot,\,UM}$) and the raw background rate density (CRD$_{\rm raw}^{\rm bkg}$).
    \item[4)] Calculate $\dot{N}_{\rm wing}$ by correcting raw count rate with Equation (\ref{equ:COI_point}-\ref{equ:N_wing}) with COI, EXT, LSS, and SEN factors.
    \item[5)] If the $\dot{N}_{\rm wing}$ is in the valid range for the PSF method, calculate the AB magnitude of the saturated source with zeropoints listed in the last column of Table. \ref{tab:ZP} and Equation (\ref{equ:AB_wing}).
\end{itemize}

For V and B bands, the maximum corrected $\dot{N}_{\rm wing}$ for PSF method is $\sim$100\,count/s, i.e. $\sim$1.7 mag brighter than saturation limit, corresponding to $\sim9.78$th mag and $\sim10.87$th mag, respectively. However, due to lack of bright U-band sources, currently the maximum corrected $\dot{N}_{\rm wing}$ for U band is only $\sim$40\,count/s, i.e. $\sim$0.7 mag brighter than saturation limit, corresponding to $\sim$12.17th mag. However, there seems pretty room to improve the range for the U band, since the saturation limit is caused by readout speed of CCD, which has nothing to do with filters. In other words, in principle, the PSF method can be applied to sources with $\dot{N}_{\rm wing}<$100\,count/s as well.

This work based on color transformation between different filter systems and indeed there exist some factors contributing to the systematic uncertainty and the bias of the PSF method. The typical value of the systematic uncertainty induced by the measuring principle is $\sim0.1$ mag, while the large scatter of converted UVOT magnitudes from other catalogues contribute to the comprehensive/equivalent systematic uncertainty mostly. Hence, well observed sources with spectra can be used to calibrate the PSF method, which would reduce the uncertainty induced by the calibration data set and possibly solve the difference found between directly measured UVOT magnitudes and converted values of nearly saturated UVOT sources (Appendix \ref{app:Difference_between_UVOT_and_JKC}).

\section{Acknowledgement}
This work was supported in part by NSFC under grants of No. 12225305, No. 11921003, No. 12233011, No. 11933010 and No. 12073080, the China Manned Space Project (NO.CMS-CSST-2021-A13), Major Science and Technology Project of Qinghai Province  (2019-ZJ-A10), Key Research Program of Frontier Sciences (No. QYZDJ-SSW-SYS024). SC has been supported by ASI grant I/004/11/0.

This research made use of Photutils, an Astropy package for detection and photometry of astronomical sources \citep{larry_bradley_2023_7946442}.

\bibliography{sample631_arxiv}{}

\begin{thebibliography}{}
\expandafter\ifx\csname natexlab\endcsname\relax\def\natexlab#1{#1}\fi
\providecommand{\url}[1]{\href{#1}{#1}}
\providecommand{\dodoi}[1]{doi:~\href{http://doi.org/#1}{\nolinkurl{#1}}}
\providecommand{\doeprint}[1]{\href{http://ascl.net/#1}{\nolinkurl{http://ascl.net/#1}}}
\providecommand{\doarXiv}[1]{\href{https://arxiv.org/abs/#1}{\nolinkurl{https://arxiv.org/abs/#1}}}

\bibitem[{{Bessell} \& {Murphy}(2012)}]{2012PASP..124..140B}
{Bessell}, M., \& {Murphy}, S. 2012, \pasp, 124, 140, \dodoi{10.1086/664083}

\bibitem[{{Bloom} {et~al.}(2009){Bloom}, {Perley}, {Li}, {Butler}, {Miller},
  {Kocevski}, {Kann}, {Foley}, {Chen}, {Filippenko}, {Starr}, {Macomber},
  {Prochaska}, {Chornock}, {Poznanski}, {Klose}, {Skrutskie}, {Lopez}, {Hall},
  {Glazebrook}, \& {Blake}}]{2009ApJ...691..723B}
{Bloom}, J.~S., {Perley}, D.~A., {Li}, W., {et~al.} 2009, \apj, 691, 723,
  \dodoi{10.1088/0004-637X/691/1/723}

\bibitem[{{Bohlin}(2014)}]{2014AJ....147..127B}
{Bohlin}, R.~C. 2014, \aj, 147, 127, \dodoi{10.1088/0004-6256/147/6/127}

\bibitem[{Bradley(2023)}]{larry_bradley_2023_7946442}
Bradley, L. 2023, astropy/photutils: 1.8.0, 1.8.0,  Zenodo,
  \dodoi{10.5281/zenodo.7946442}

\bibitem[{{Breeveld} {et~al.}(2011){Breeveld}, {Landsman}, {Holland}, {Roming},
  {Kuin}, \& {Page}}]{2011AIPC.1358..373B}
{Breeveld}, A.~A., {Landsman}, W., {Holland}, S.~T., {et~al.} 2011, in American
  Institute of Physics Conference Series, Vol. 1358, Gamma Ray Bursts 2010, ed.
  J.~E. {McEnery}, J.~L. {Racusin}, \& N.~{Gehrels}, 373--376,
  \dodoi{10.1063/1.3621807}

\bibitem[{{Breeveld} {et~al.}(2010){Breeveld}, {Curran}, {Hoversten}, {Koch},
  {Landsman}, {Marshall}, {Page}, {Poole}, {Roming}, {Smith}, {Still},
  {Yershov}, {Blustin}, {Brown}, {Gronwall}, {Holland}, {Kuin}, {McGowan},
  {Rosen}, {Boyd}, {Broos}, {Carter}, {Chester}, {Hancock}, {Huckle}, {Immler},
  {Ivanushkina}, {Kennedy}, {Mason}, {Morgan}, {Oates}, {de Pasquale},
  {Schady}, {Siegel}, \& {vanden Berk}}]{2010MNRAS.406.1687B}
{Breeveld}, A.~A., {Curran}, P.~A., {Hoversten}, E.~A., {et~al.} 2010, \mnras,
  406, 1687, \dodoi{10.1111/j.1365-2966.2010.16832.x}

\bibitem[{{Esa}(1997)}]{1997yCat.1239....0E}
{Esa}, . 1997, VizieR Online Data Catalog, I/239

\bibitem[{{Fordham} {et~al.}(2000){Fordham}, {Moorhead}, \&
  {Galbraith}}]{2000MNRAS.312...83F}
{Fordham}, J.~L.~A., {Moorhead}, C.~F., \& {Galbraith}, R.~F. 2000, \mnras,
  312, 83, \dodoi{10.1046/j.1365-8711.2000.03155.x}

\bibitem[{{Gaia Collaboration} {et~al.}(2022){Gaia Collaboration},
  {Montegriffo}, {Bellazzini}, {De Angeli}, {Andrae}, {Barstow}, {Bossini},
  {Bragaglia}, {Burgess}, {Cacciari}, {Carrasco}, {Chornay}, {Delchambre},
  {Evans}, {Fouesneau}, {Fremat}, {Garabato}, {Jordi}, {Manteiga}, {Massari},
  {Palaversa}, {Pancino}, {Riello}, {Ruz Mieres}, {Sanna}, {Santovena},
  {Sordo}, {Vallenari}, {Walton}, \& {DPAC}}]{2022arXiv220606215G}
{Gaia Collaboration}, {Montegriffo}, P., {Bellazzini}, M., {et~al.} 2022, arXiv
  e-prints, arXiv:2206.06215.
\newblock \doarXiv{2206.06215}

\bibitem[{{H{\o}g} {et~al.}(2000){H{\o}g}, {Fabricius}, {Makarov}, {Urban},
  {Corbin}, {Wycoff}, {Bastian}, {Schwekendiek}, \&
  {Wicenec}}]{2000A&A...355L..27H}
{H{\o}g}, E., {Fabricius}, C., {Makarov}, V.~V., {et~al.} 2000, \aap, 355, L27

\bibitem[{{Jin} {et~al.}(2023){Jin}, {Zhou}, {Wang}, {Geng}, {Covino}, {Wu},
  {Li}, {Fan}, {Wei}, \& {Wei}}]{2023NatAs.tmp..135J}
{Jin}, Z.-P., {Zhou}, H., {Wang}, Y., {et~al.} 2023, Nature Astronomy,
  \dodoi{10.1038/s41550-023-02005-w}

\bibitem[{{Kuin} {et~al.}(2021){Kuin}, {Lien}, \& {Swift/UVOT
  Team}}]{2021GCN.30356....1K}
{Kuin}, N.~P.~M., {Lien}, A.~Y., \& {Swift/UVOT Team}. 2021, GRB Coordinates
  Network, 30356, 1

\bibitem[{{Kuin} {et~al.}(2022){Kuin}, {Tohuvavohu}, \& {Swift/UVOT
  Team}}]{2022GCN.31351....1K}
{Kuin}, N.~P.~M., {Tohuvavohu}, A., \& {Swift/UVOT Team}. 2022, GRB Coordinates
  Network, 31351, 1

\bibitem[{{Kuin} {et~al.}(2009){Kuin}, {Landsman}, {Page}, {Schady}, {Still},
  {Breeveld}, {de Pasquale}, {Roming}, {Brown}, {Carter}, {James}, {Curran},
  {Cucchiara}, {Gronwall}, {Holland}, {Hoversten}, {Hunsberger}, {Kennedy},
  {Koch}, {Lamoureux}, {Marshall}, {Oates}, {Parsons}, {Palmer}, \&
  {Smith}}]{2009MNRAS.395L..21K}
{Kuin}, N.~P.~M., {Landsman}, W., {Page}, M.~J., {et~al.} 2009, \mnras, 395,
  L21, \dodoi{10.1111/j.1745-3933.2009.00632.x}

\bibitem[{{Kuin} {et~al.}(2015){Kuin}, {Landsman}, {Breeveld}, {Page},
  {Lamoureux}, {James}, {Mehdipour}, {Still}, {Yershov}, {Brown}, {Carter},
  {Mason}, {Kennedy}, {Marshall}, {Roming}, {Siegel}, {Oates}, {Smith}, \& {De
  Pasquale}}]{2015MNRAS.449.2514K}
{Kuin}, N.~P.~M., {Landsman}, W., {Breeveld}, A.~A., {et~al.} 2015, \mnras,
  449, 2514, \dodoi{10.1093/mnras/stv408}

\bibitem[{{Ma{\'\i}z-Apell{\'a}niz}(2005)}]{2005PASP..117..615M}
{Ma{\'\i}z-Apell{\'a}niz}, J. 2005, \pasp, 117, 615, \dodoi{10.1086/430370}

\bibitem[{{Margutti} {et~al.}(2014){Margutti}, {Parrent}, {Kamble},
  {Soderberg}, {Foley}, {Milisavljevic}, {Drout}, \&
  {Kirshner}}]{2014ApJ...790...52M}
{Margutti}, R., {Parrent}, J., {Kamble}, A., {et~al.} 2014, \apj, 790, 52,
  \dodoi{10.1088/0004-637X/790/1/52}

\bibitem[{{Maselli} {et~al.}(2014){Maselli}, {Melandri}, {Nava}, {Mundell},
  {Kawai}, {Campana}, {Covino}, {Cummings}, {Cusumano}, {Evans}, {Ghirlanda},
  {Ghisellini}, {Guidorzi}, {Kobayashi}, {Kuin}, {La Parola}, {Mangano},
  {Oates}, {Sakamoto}, {Serino}, {Virgili}, {Zhang}, {Barthelmy}, {Beardmore},
  {Bernardini}, {Bersier}, {Burrows}, {Calderone}, {Capalbi}, {Chiang},
  {D'Avanzo}, {D'Elia}, {De Pasquale}, {Fugazza}, {Gehrels}, {Gomboc},
  {Harrison}, {Hanayama}, {Japelj}, {Kennea}, {Kopac}, {Kouveliotou}, {Kuroda},
  {Levan}, {Malesani}, {Marshall}, {Nousek}, {O'Brien}, {Osborne}, {Pagani},
  {Page}, {Page}, {Perri}, {Pritchard}, {Romano}, {Saito}, {Sbarufatti},
  {Salvaterra}, {Steele}, {Tanvir}, {Vianello}, {Weigand}, {Wiersema}, {Yatsu},
  {Yoshii}, \& {Tagliaferri}}]{2014Sci...343...48M}
{Maselli}, A., {Melandri}, A., {Nava}, L., {et~al.} 2014, Science, 343, 48,
  \dodoi{10.1126/science.1242279}

\bibitem[{{Page} {et~al.}(2013){Page}, {Kuin}, {Breeveld}, {Hancock},
  {Holland}, {Marshall}, {Oates}, {Roming}, {Siegel}, {Smith}, {Carter}, {De
  Pasquale}, {Symeonidis}, {Yershov}, \& {Beardmore}}]{2013MNRAS.436.1684P}
{Page}, M.~J., {Kuin}, N.~P.~M., {Breeveld}, A.~A., {et~al.} 2013, \mnras, 436,
  1684, \dodoi{10.1093/mnras/stt1689}

\bibitem[{{Page} {et~al.}(2014){Page}, {Yershov}, {Breeveld}, {Kuin},
  {Mignani}, {Smith}, {Rawlings}, {Oates}, {Siegel}, \&
  {Roming}}]{2014styd.confE..37P}
{Page}, M.~J., {Yershov}, V., {Breeveld}, A., {et~al.} 2014, in Proceedings of
  Swift: 10 Years of Discovery (SWIFT 10, 37.
\newblock \doarXiv{1503.06597}

\bibitem[{{Pickles}(1998)}]{1998PASP..110..863P}
{Pickles}, A.~J. 1998, \pasp, 110, 863, \dodoi{10.1086/316197}

\bibitem[{{Poole} {et~al.}(2008){Poole}, {Breeveld}, {Page}, {Landsman},
  {Holland}, {Roming}, {Kuin}, {Brown}, {Gronwall}, {Hunsberger}, {Koch},
  {Mason}, {Schady}, {vanden Berk}, {Blustin}, {Boyd}, {Broos}, {Carter},
  {Chester}, {Cucchiara}, {Hancock}, {Huckle}, {Immler}, {Ivanushkina},
  {Kennedy}, {Marshall}, {Morgan}, {Pandey}, {de Pasquale}, {Smith}, \&
  {Still}}]{2008MNRAS.383..627P}
{Poole}, T.~S., {Breeveld}, A.~A., {Page}, M.~J., {et~al.} 2008, \mnras, 383,
  627, \dodoi{10.1111/j.1365-2966.2007.12563.x}

\bibitem[{{Racusin} {et~al.}(2008){Racusin}, {Karpov}, {Sokolowski}, {Granot},
  {Wu}, {Pal'Shin}, {Covino}, {van der Horst}, {Oates}, {Schady}, {Smith},
  {Cummings}, {Starling}, {Piotrowski}, {Zhang}, {Evans}, {Holland}, {Malek},
  {Page}, {Vetere}, {Margutti}, {Guidorzi}, {Kamble}, {Curran}, {Beardmore},
  {Kouveliotou}, {Mankiewicz}, {Melandri}, {O'Brien}, {Page}, {Piran},
  {Tanvir}, {Wrochna}, {Aptekar}, {Barthelmy}, {Bartolini}, {Beskin}, {Bondar},
  {Bremer}, {Campana}, {Castro-Tirado}, {Cucchiara}, {Cwiok}, {D'Avanzo},
  {D'Elia}, {Della Valle}, {de Ugarte Postigo}, {Dominik}, {Falcone}, {Fiore},
  {Fox}, {Frederiks}, {Fruchter}, {Fugazza}, {Garrett}, {Gehrels},
  {Golenetskii}, {Gomboc}, {Gorosabel}, {Greco}, {Guarnieri}, {Immler},
  {Jelinek}, {Kasprowicz}, {La Parola}, {Levan}, {Mangano}, {Mazets},
  {Molinari}, {Moretti}, {Nawrocki}, {Oleynik}, {Osborne}, {Pagani}, {Pandey},
  {Paragi}, {Perri}, {Piccioni}, {Ramirez-Ruiz}, {Roming}, {Steele}, {Strom},
  {Testa}, {Tosti}, {Ulanov}, {Wiersema}, {Wijers}, {Winters}, {Zarnecki},
  {Zerbi}, {M{\'e}sz{\'a}ros}, {Chincarini}, \&
  {Burrows}}]{2008Natur.455..183R}
{Racusin}, J.~L., {Karpov}, S.~V., {Sokolowski}, M., {et~al.} 2008, \nat, 455,
  183, \dodoi{10.1038/nature07270}

\bibitem[{{Roming} {et~al.}(2005){Roming}, {Kennedy}, {Mason}, {Nousek}, {Ahr},
  {Bingham}, {Broos}, {Carter}, {Hancock}, {Huckle}, {Hunsberger}, {Kawakami},
  {Killough}, {Koch}, {McLelland}, {Smith}, {Smith}, {Soto}, {Boyd},
  {Breeveld}, {Holland}, {Ivanushkina}, {Pryzby}, {Still}, \&
  {Stock}}]{2005SSRv..120...95R}
{Roming}, P. W.~A., {Kennedy}, T.~E., {Mason}, K.~O., {et~al.} 2005, \ssr, 120,
  95, \dodoi{10.1007/s11214-005-5095-4}

\bibitem[{{Su} {et~al.}(2022){Su}, {Rieke}, {Marengo}, \&
  {Schlawin}}]{2022AJ....163...46S}
{Su}, K. Y.~L., {Rieke}, G.~H., {Marengo}, M., \& {Schlawin}, E. 2022, \aj,
  163, 46, \dodoi{10.3847/1538-3881/ac3b5e}

\bibitem[{{Wo{\'z}niak} {et~al.}(2009){Wo{\'z}niak}, {Vestrand}, {Panaitescu},
  {Wren}, {Davis}, \& {White}}]{2009ApJ...691..495W}
{Wo{\'z}niak}, P.~R., {Vestrand}, W.~T., {Panaitescu}, A.~D., {et~al.} 2009,
  \apj, 691, 495, \dodoi{10.1088/0004-637X/691/1/495}

\end{thebibliography}
\bibliographystyle{aasjournal}

\clearpage
\begin{table}[ht]
    \centering
    \begin{tabular}{ccc}
        \hline
        \hline
        Filter & Valid $\dot{N}_{\rm wing}$ & $ZP^{\rm AB}_{\rm wing}$ \\
        & (count/s) & \\
        \hline
        V & 10\,-\,100 & $14.774\pm0.182$ \\
        B & 20\,-\,100 & $15.872\pm0.178$ \\
        U & 12\,-\,40 & $16.177\pm0.165$ \\
        \hline
    \end{tabular} \\
    \caption{Zeropoints of PSF method. The first column represents names of UVOT filters. The second column lists the valid range of $\dot{N}_{\rm wing}$ for each band. The last column is the zeropoint calibrated with all data sets for a specific band. Differences between zeropoints for different operation and bin modes are very small ($<\sim0.02$ mag, please see Table \ref{tab:ZP_Individual}), hence it is safe to use these values as the zeropoint of the PSF method for a specified band. All values following ``$\pm$" are 1-$\sigma$ uncertainties of zeropoints, i.e., the systematic uncertainty of the PSF method.}
    \label{tab:ZP}
\end{table}

\begin{table}[h]
    \centering
    \begin{tabular}{ccccccc}
        \hline
        \hline
        T$_{\rm mid}$-T$_0$ & Exp & $\dot{N}_{\rm wing}$ & SNR & V$_{\rm wing}$\,$^{a}$ & V$_{\rm ref}$\,$^{b}$ & V$_{\it uvotsource}$\,$^{c}$\\
        (s) & (s) & (count/s) & & (AB) & (Vega) & (AB) \\
        \hline
        \hline
        \multicolumn{7}{c}{Time bins of \cite{2009ApJ...691..495W}} \\
        \hline
        180.60 & 9.84 & 69.61$\pm$3.29 & 20.53 & 10.17$\pm$0.19 & 10.06$\pm$0.02 & / \\
        193.52 & 9.84 & 57.48$\pm$3.05 & 18.40 & 10.38$\pm$0.20 & 10.23$\pm$0.02 & / \\
        206.25 & 9.84 & 44.96$\pm$2.78 & 15.72 & 10.64$\pm$0.20 & 10.44$\pm$0.02 & / \\
        218.97 & 9.84 & 42.13$\pm$2.71 & 15.16 & 10.71$\pm$0.20 & 10.55$\pm$0.02 & / \\
        231.70 & 9.84 & 37.33$\pm$2.61 & 13.65 & 10.84$\pm$0.20 & 10.72$\pm$0.02 & / \\
        244.43 & 9.84 & 31.39$\pm$2.47 & 12.47 & 11.03$\pm$0.21 & 10.88$\pm$0.02 & / \\
        257.45 & 9.84 & 28.22$\pm$2.39 & 11.33 & 11.15$\pm$0.21 & 11.00$\pm$0.02 & / \\
        270.38 & 9.84 & 26.66$\pm$2.36 & 11.25 & 11.21$\pm$0.21 & 11.12$\pm$0.02 & / \\
        283.30 & 9.84 & 24.81$\pm$2.30 & 10.52 & 11.29$\pm$0.21 & 11.23$\pm$0.02 & 11.42$\pm$0.07 \\
        296.23 & 9.84 & 23.21$\pm$2.25 & 9.98 & 11.36$\pm$0.22 & 11.36$\pm$0.02 & 11.54$\pm$0.07 \\
        309.15 & 9.84 & 18.61$\pm$2.13 & 8.44 & 11.60$\pm$0.23 & 11.52$\pm$0.02 & 11.70$\pm$0.06$^d$ \\
        322.08 & 9.84 & 14.57$\pm$2.05 & 6.85 & 11.87$\pm$0.24 & 11.61$\pm$0.02 & 11.70$\pm$0.06$^d$ \\
        334.80 & 9.84 & 15.24$\pm$2.06 & 7.60 & 11.82$\pm$0.24 & 11.72$\pm$0.02 & 11.79$\pm$0.06 \\
        360.05 & 29.52 & 11.39$\pm$1.11 & 9.91 & 12.13$\pm$0.22 & 11.91$\pm$0.02 & 11.98$\pm$0.04 \\
        395.50 & 29.52 & 10.34$\pm$1.09 & 9.11 & 12.24$\pm$0.22 & 12.20$\pm$0.02 & 12.20$\pm$0.04 \\
        431.04 & 29.52 & 9.06$\pm$1.07 & 8.07 & 12.38$\pm$0.23 & 12.40$\pm$0.02 & 12.42$\pm$0.04 \\
        466.29 & 29.52 & 7.95$\pm$1.05 & 7.07 & 12.52$\pm$0.24 & 12.54$\pm$0.02 & 12.53$\pm$0.04 \\
        502.44 & 29.52 & 6.56$\pm$1.02 & 6.08 & 12.73$\pm$0.25 & 12.74$\pm$0.02 & 12.71$\pm$0.04 \\
        537.68 & 29.52 & 5.60$\pm$1.00 & 5.28 & 12.90$\pm$0.27 & 12.80$\pm$0.02 & 12.86$\pm$0.04 \\
        \hline
        \hline
        \multicolumn{7}{c}{Time bins of \cite{2013MNRAS.436.1684P}} \\
        \hline
        189.92 & 29.49 & 60.25$\pm$1.78 & 33.04 & 10.32$\pm$0.19 & 10.07$_{-0.22}^{+0.26}$ & / \\
        224.89 & 39.36 & 39.14$\pm$1.29 & 29.64 & 10.79$\pm$0.19 & 10.44$_{-0.24}^{+0.29}$ & / \\
        269.89 & 49.21 & 26.80$\pm$1.02 & 25.60 & 11.20$\pm$0.19 & 10.89$_{-0.29}^{+0.38}$ & / \\
        322.39 & 54.13 & 17.95$\pm$0.87 & 20.12 & 11.64$\pm$0.19 & 11.60$_{-0.45}^{+0.74}$ & 11.75$\pm$0.02 \\
        \hline
    \end{tabular} \\
    $a$. Calculated with $\dot{N}_{\rm wing}$ with the zorepoint for V band. \\
    $b$. These values are directly taken from \cite{2009ApJ...691..495W} and \cite{2013MNRAS.436.1684P} and in Vega system. \\
    $c$. Measured with UVOT FTOOLS command {\it uvotsource}. Saturated time bins are marked with ``/". \\
    $d$. More accurate values are 11.69511 and 11.70301.
    \caption{V-band measurements of GRB 080319B. All data points have been plotted in Figure \ref{fig:GRB_lc} for a comparison, except the photomtry measured with $uvotsource$ for the last time bin of \cite{2013MNRAS.436.1684P}.}
    \label{tab:080319B_V}
\end{table}

\newpage

\newpage
\clearpage

\begin{figure}[h]
    \centering
    \includegraphics[width=0.5\columnwidth]{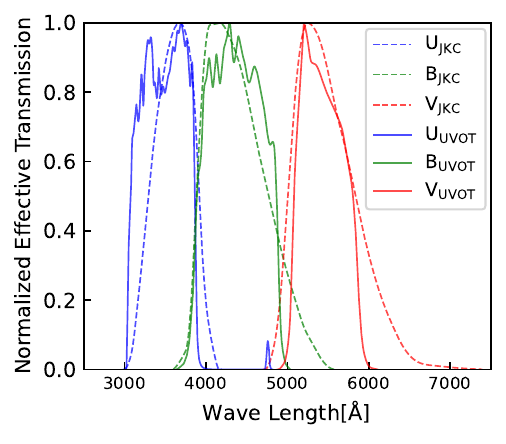}
    \caption{Normalized transmission curves of Jonhnson-Kron-Cousions filters and UVOT filters. Dashed lines are normalized transmission curve of Johnson-Kron-Cousins filters and solid lines represent that of UVOT filters. Blue, green and red lines represent U, B and V bands, respectively. The effect of quantum efficiency of CCD is applied to all curves.}
    \label{fig:VBU_Comparison}
\end{figure}

\newpage
\clearpage
\begin{figure}[ht]
    \centering
    \includegraphics[width=0.32\columnwidth]{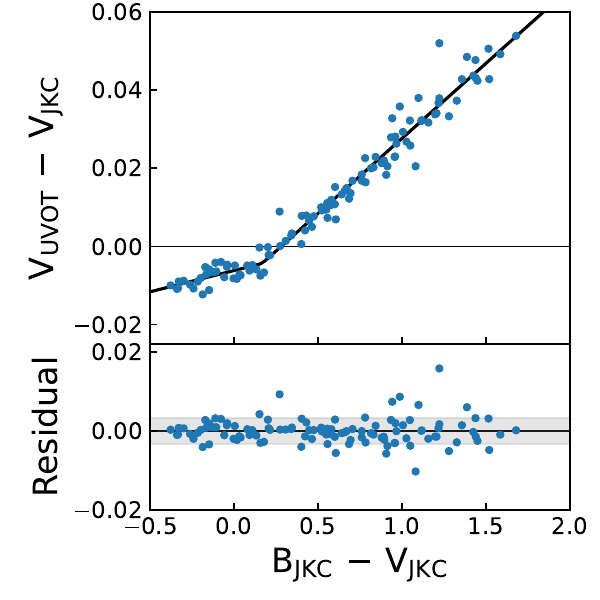}
    \includegraphics[width=0.32\columnwidth]{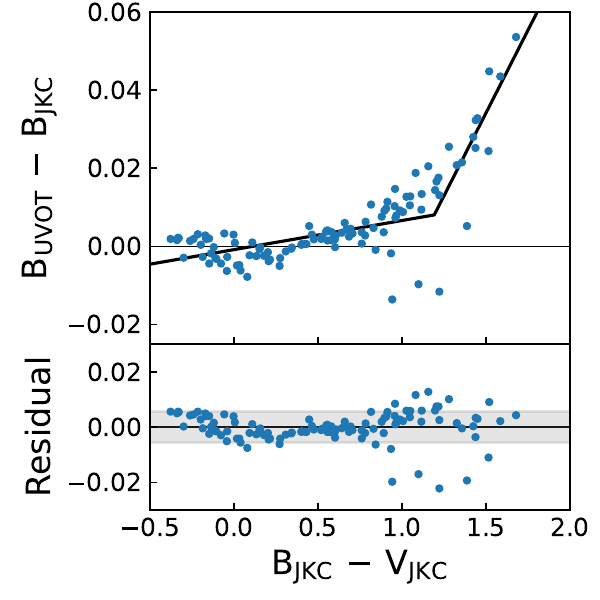}
    \includegraphics[width=0.32\columnwidth]{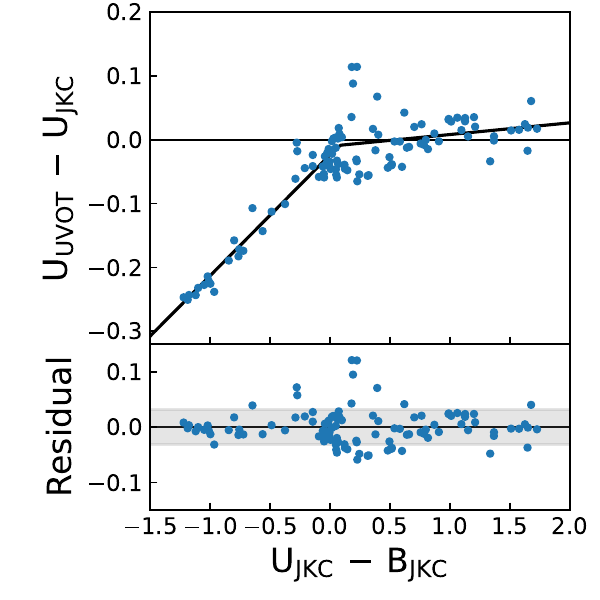}
    \caption{Color transformation from Johnson-Kron-Cousin system to UVOT system. From left to right, 3 panels show color transformation of V, B and U bands, respectively. Since main peaks of V and B bands of Jhonson-Kron-Cousins system and UVOT system are very close(see Figure \ref{fig:VBU_Comparison}), ${\rm B_{UVOT} - B_{JKC}}$ and ${\rm V_{UVOT} - V_{JKC}}$ are very close to 0, i.e. $< 0.06$ mag, as shown in the figure. For U band, the difference between 2 systems reaches a maximum of $\sim$ 0.25 mag. Note that in this figure, color terms are under Vega system, and AB magnitudes of Vega in UVOT V, B, and U bands are -0.01, -0.13, and 1.02 mag, respectively. M type stars are not plotted in this figure, due to their large scatter.}
    \label{fig:color_transformation}
\end{figure}

\newpage
\clearpage
\begin{figure}[ht]
    \centering
    \includegraphics[width=\columnwidth]{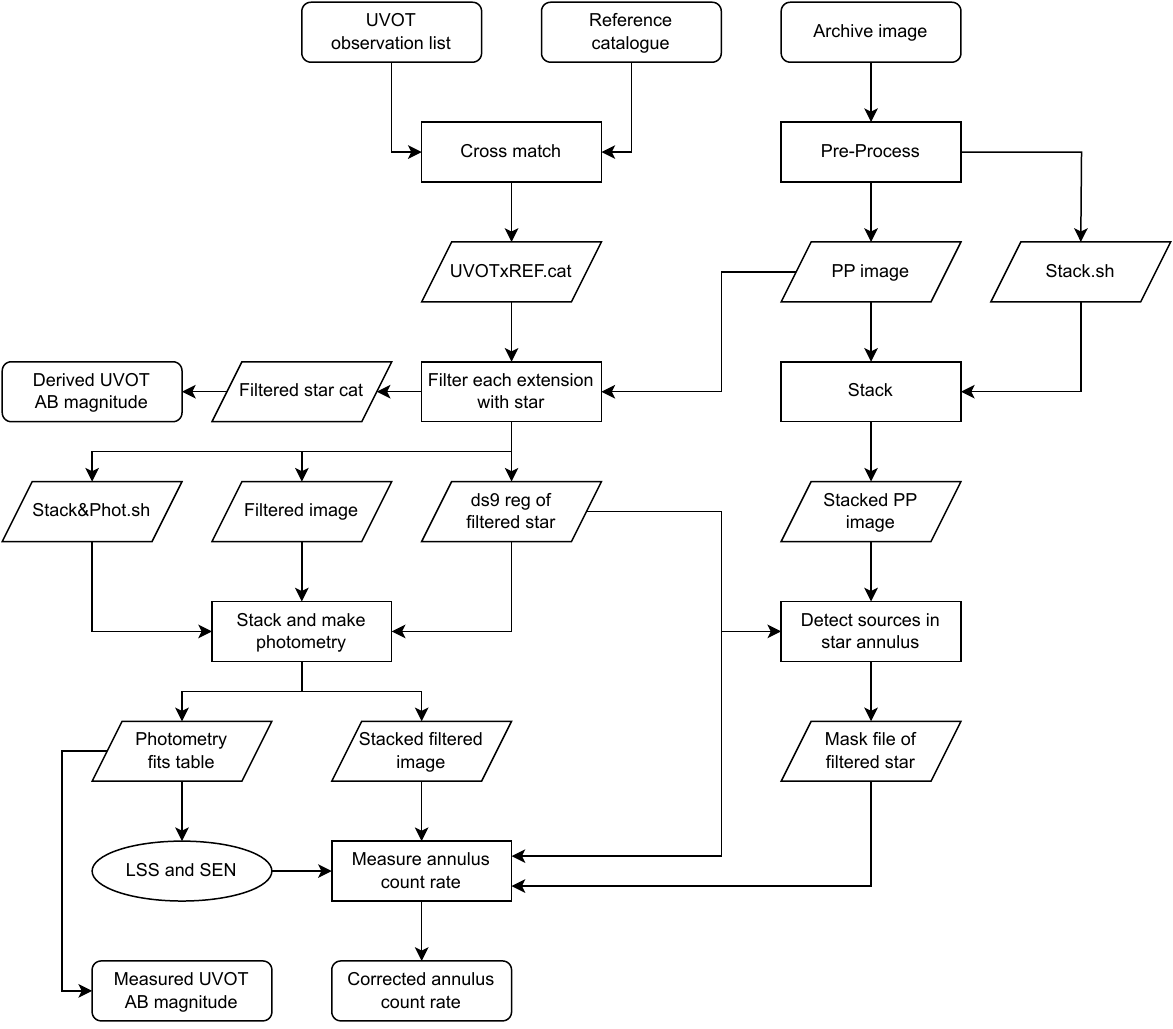}
    \caption{The flow chart for the procedures described in Section \ref{subsec:generation_of_calibration_data}. A simple summary: 1) The UVOT observation list is cross matched with the reference catalogue with a search radius of 12\,arcmin. 2) Each extension of PP images is checked with the corresponding matched star, because the pointing coordinate and the field of view of each extension could be different though the extensions are in one exposure sequence. 3) Measure the count rate of the wing for each star that passed all selection criteria.}
    \label{fig:pipeline}
\end{figure}

\newpage
\clearpage
\begin{figure}[ht]
    \centering
    \includegraphics[width=\columnwidth]{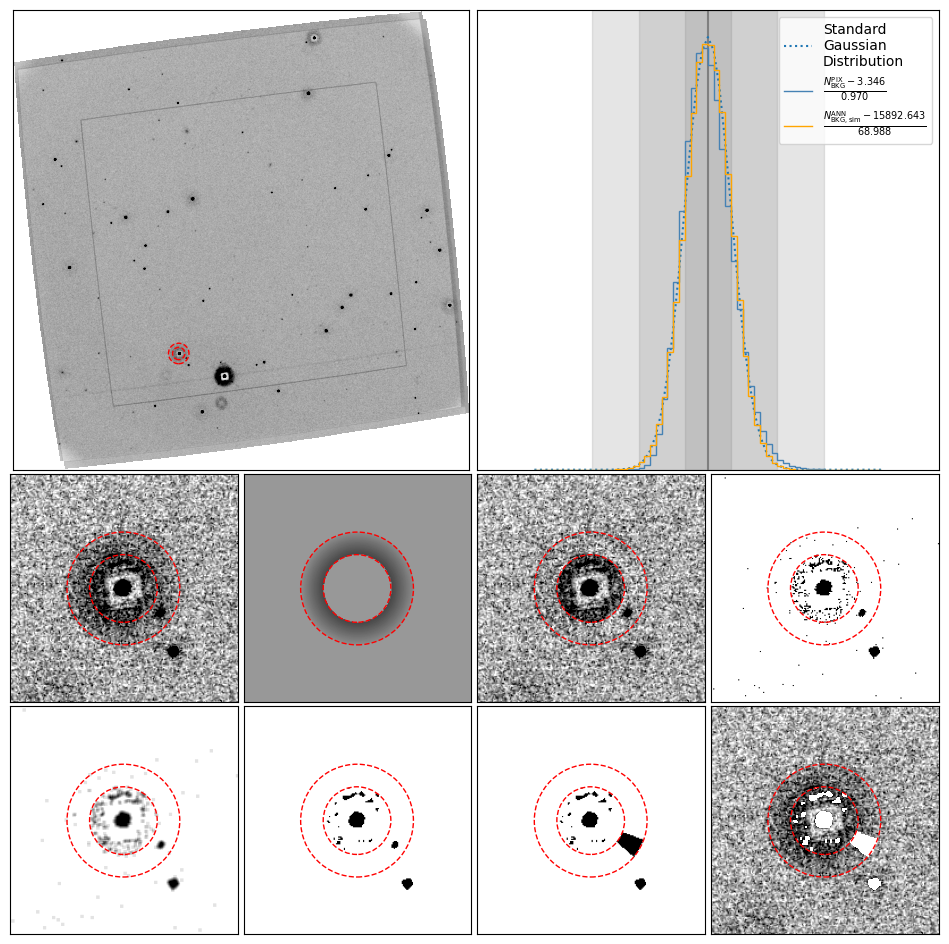}
    \caption{An example for measurement of the wing. This figure is generated by the algorithm to measure the count rate of the wing. The 2 panels at the top row are normalized sky image and the statistical information for the background. For the mid row, the 4 panels from left to right are the magnified image of the source, the fitted model for the wing, the residual image for the wing, and the bad pixel map created with the sigma clip method with a threshold of 5. It is clearly to be seen there are some true background pixels are marked as bad pixels and they should be removed from the bad pixel map to avoid their influence on the generation of the mask for the wing. For the bottom row, 4 panels from left to right are the smoothed bad pixel map, the clean bad pixel map, the mask for the wing, and the masked magnified image of the wing. Please note that the mask for sources in the wing should be composed with sector annuli instead of source shapes, or the Equation (\ref{equ:COI_wing}) is wrong.} 
    \label{fig:example_measurement}
\end{figure}

\newpage
\clearpage
\begin{figure}[ht]
    \centering
    \includegraphics[width=0.35\columnwidth]{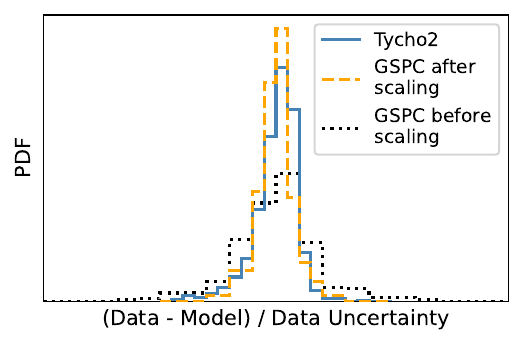}
    \caption{An illustration of the uncertainty weighted residual distribution for the V-band event 1x1 data set. The GSPC data will have higher weight than Tycho-2 data. With the scaling factor, GSPC and Tycho-2 have approximate influence on the fitting for the calibration of zeropoints.} 
    \label{fig:V_res_dis}
\end{figure}

\begin{figure}[ht]
    \centering
    \includegraphics[width=0.4\columnwidth]{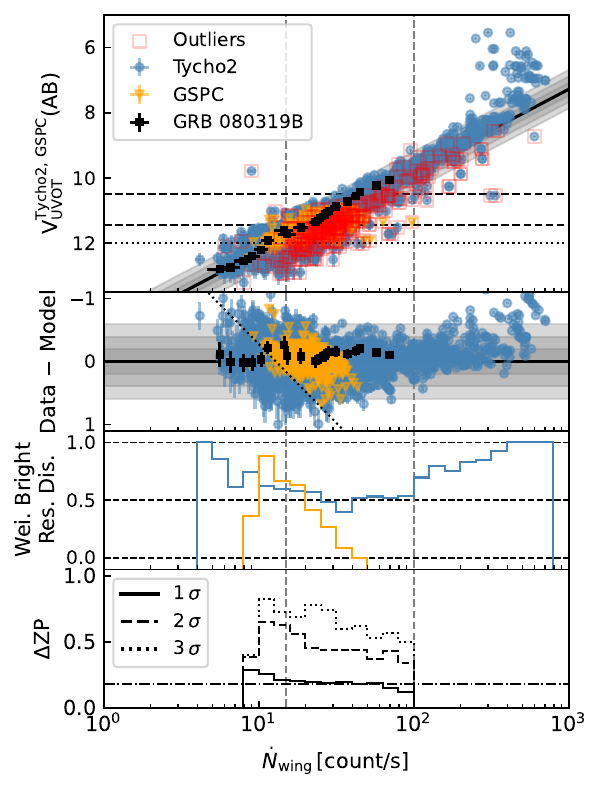}
    \caption{The V-band calibration for the PSF method. The first panel represents the result for the calibration. Blue and orange points are Tycho-2 and GSPC data points, respectively. Outliers for the fitting are marked with a red square. The 2 vertical dashed lines mark the range of Tycho-2 data points used for the fitting procedure. The lower horizontal dashed represents the COI saturation limit at a zero background. While the upper one is for the readout streak method: for sources brighter than this value, the uncertainty derived with the readout streak method is comparable or smaller than the typical value $\sim20\%$ of the PSF method. The horizontal dotted line marks the lower limit for GSPC data points, which the theoretical residual of this dotted line is shown in the second panel to illustrate that the strange shape of GSPC residuals is not induced by any physical reasons. The gray region represents 1, 2, and 3 times the RMS of residuals, which is also shown in the second panel. For V band, measurements of GRB 080319B, labeled with black points, are plotted but not fitted. Fitting residuals are shown in the second panel, where outliers are not plotted for clarity. The third panel represents the weighted bright residual distribution. The vertical axis represents the fraction of sources brighter than the fitted model weighted by reciprocals of uncertainties of data points. The domination of bright residuals at low count rate is due to the lower limit in photometries of Tycho-2 sources. While for high count rate, the wing strongly influenced by the coincidence loss. The last panel represents the systematic uncertainty for the PSF method. Solid, dashed, and dotted lines represent 1$\,\sigma$, 2$\,\sigma$, and 3$\,\sigma$ uncertainties for each bin, respectively. The dot-dashed line is the mean value of 1$\,\sigma$ uncertainties, please see Table \ref{tab:ZP} for exact values.} 
    \label{fig:V_all}
\end{figure}

\newpage
\clearpage
\begin{figure}[ht]
    \centering
    \includegraphics[width=0.4\columnwidth]{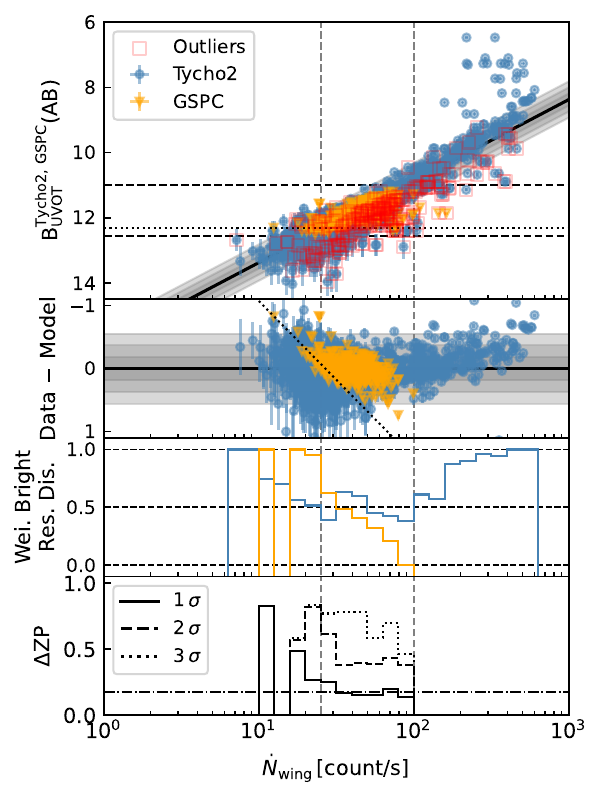}
    \caption{The B-band calibration for the PSF method. Please refer to the caption of Figure \ref{fig:V_all} for the meaning of all notations.} 
    \label{fig:B_all}
\end{figure}

\begin{figure}[h]
    \centering
    \includegraphics[width=0.4\columnwidth]{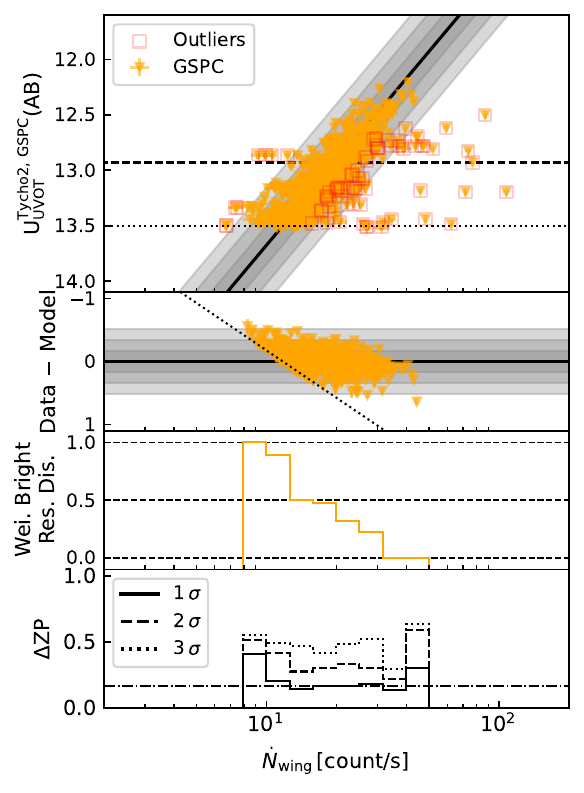}
    \caption{The U-band calibration for the PSF method. Please refer to the caption of Figure \ref{fig:V_all} for the meaning of all notations. Except the only dashed line in the first panel represents the COI saturation limit of UVOT.} 
    \label{fig:U_all}
\end{figure}

\newpage
\clearpage
\begin{figure}[ht]
    \centering
    \includegraphics[width=0.4\columnwidth]{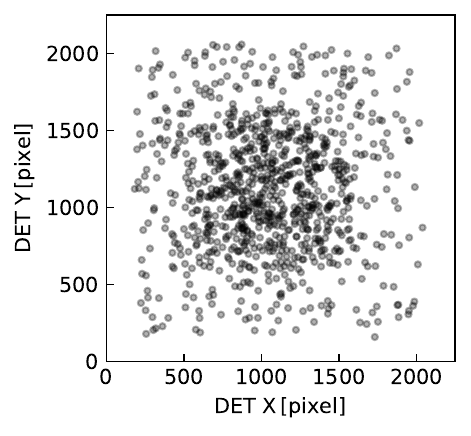}
    \caption{The spatial distribution of calibration sources in V-band event 1x1 data set. This figure shows the detector position of each sources used for the calibration of $ZP_{\rm wing,\,evt1x1}^{\rm AB,\,V}$. Points with darker color are the result that several individual points coincides with each other.}
    \label{fig:cal_src_dis_V_evt1x1}
\end{figure}

\begin{figure}[h]
    \centering
    \includegraphics[width=0.7\columnwidth]{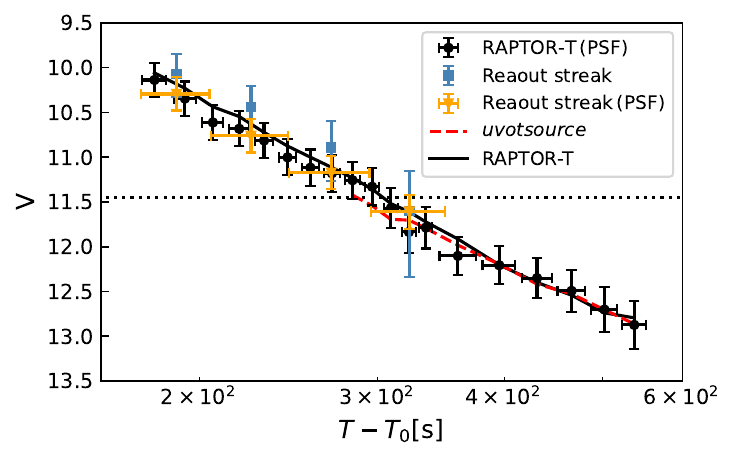}
    \caption{The light curve of GRB 080319B in V band. The black solid line represents RAPTOR-T measurements \citep[taken from][]{2009ApJ...691..495W}. The red dashed line is measurements obtained with UVOT FTOOLS $uvotsource$. Because uncertainties of measurements of RAPTOR-T and $uvotsource$ are not plotted for clarity, because compared with other points, they are very small $<0.07$ mag. Black points show the photometreis restored by the PSF method with same time bins of \cite{2009ApJ...691..495W}. Blue points show restored photometries by the readout streak method \citep{2013MNRAS.436.1684P} and photometries retored by the PSF method with same time bins of \cite{2013MNRAS.436.1684P} are plotted with orange points (for clarity, time uncertainties of blue points are not plotted since they are same as that of orange points).}
    \label{fig:GRB_lc}
\end{figure}

\newpage
\begin{appendix}
\section{The effect of the intrinsic non-uniformity of the wing}
\label{app:NU}
The spatial distribution of photons in the wing is not uniform. Hence, it is necessary to consider the effect of non-uniformity.
However, the coincidence loss is an area effect, which means a sufficient number of pixels need to be considered to make appropriate correction.
As a result, only a theoretical maximum value for the Non-Uniformity (NU) effect can be provided as a reference.

\begin{figure}[h]
    \centering
    \includegraphics[width=0.6\columnwidth]{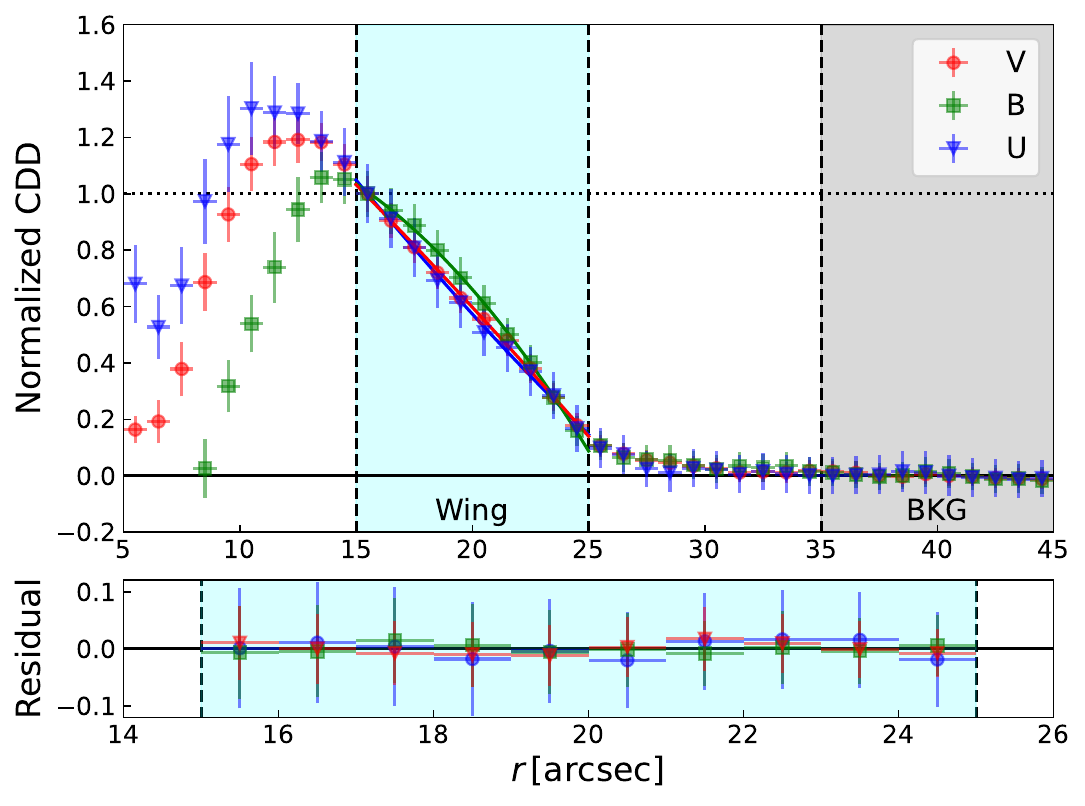}
    \caption{The fitted template of the count density distribution. Red, green and blue points/lines represent extracted CDD/fitted templates for V, B and U bands, respectively. The cyan region region represents the wing of the PSF and the background is estimated from the gray region. For the region with $r<15\,$arcsec, the coincidence loss strongly affected the PSF.} 
    \label{fig:CDD}
\end{figure}

The template of the Count Density Distribution (CDD), which represents the counts per unit area at a specific radius, is derived by stacking images of $\sim$30 isolated stars for V, B, and U bands. 
The CDD template is fitted with Chebyshev polynomials of 0th, 1st, and 2nd orders. The [0th, 1st, 2nd] coefficients for V, B, and U bands are: [2.193856, -0.070483, -0.000234], [0.965346, 0.064040, -0.001984], and [2.756965, -0.128014, 0.000472]. The fitting result for the CDD is presented in Figure \ref{fig:CDD}.

\begin{figure}[ht]
    \centering
    \includegraphics[width=0.85\columnwidth]{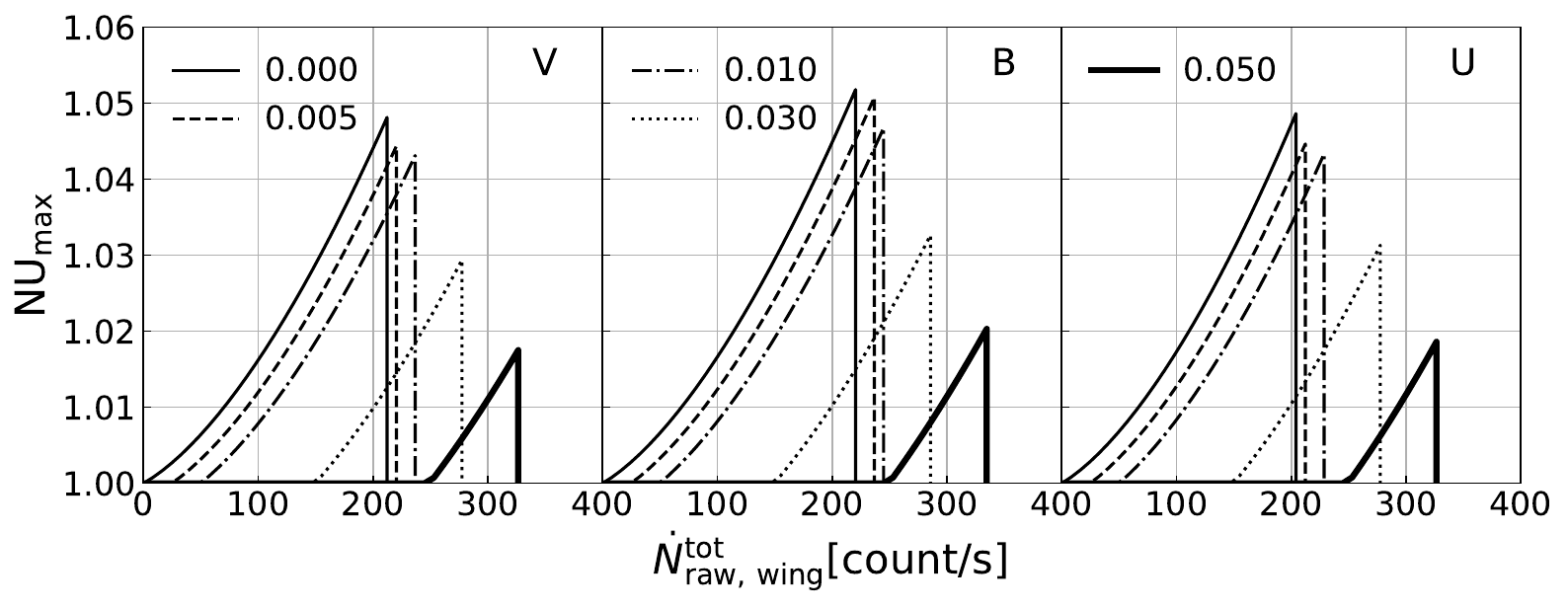}
    \caption{The maximum influence of non-uniformity. From left to right, 3 sub figures are for V, B and U bands (labels at the upper right corner of each sub figure). Thin solid, dashed, dot-dashed, dotted, and thick solid lines in all 3 sub figures represent the NU$_{\rm max}$ calculated with the raw background count rate densities CRD$_{\rm raw}^{\rm bkg}$ are 0, 0.005, 0.01, 0.03 and 0.05\,count/s/pixel, respectively. The rapid drop at high rates is because the scaled count rate used to calculate the COI and the EXT factors for the inner edge of the wing is larger than valid threshold for the fitted model of the EXT factor.} 
    \label{fig:NU_factor}
\end{figure}

Wing models with different count rates at a specific background level can be generated with CDD templates.
Then, the entire wing is divided into infinitely narrow annuli with area of $2\pi rdr$, where $r$ is the radius of the narrow annulus and $dr$ is the infinitely small width. The COI and EXT corrected total count rate of the infinitely narrow annulus with a radius of $r$, denoted as $d\dot{N}_{\rm CE,\,ann}^{\rm tot}(r)$ (the subscript ``CE" is the combination of first capitals for COI and EXT, indicating corrections that have been applied), can be calculated as:
\begin{equation}
    \begin{aligned}
        \dot{N}_{\rm raw, wing}^{\rm tot}&=A\int_{\rm 15[arcsec]}^{\rm 25[arcsec]}{\rm CDD}(r)\,2\pi rdr\,+\,400\pi[{\rm arcsec}^2]\,{\rm CRD}_{\rm raw}^{\rm bkg}\\
        \dot{N}_{\rm raw,\,ann}^{\rm COI,\,tot}(r)&=(A\,{\rm CDD}(r)\,+\,{\rm CRD}_{\rm raw}^{\rm bkg})\times25\pi[{\rm arcsec}^2] \\
        d\dot{N}_{\rm CE,\,ann}^{\rm tot}(r)&=(A\,{\rm CDD}(r)\,+\,{\rm CRD}_{\rm raw}^{\rm bkg})\,{\rm COI}_{\rm raw,\,ann}^{\rm tot}(r)\,{\rm EXT}_{\rm raw,\,ann}^{\rm tot}(r)\times2\pi rdr
        \label{equ:dNU}
    \end{aligned}
\end{equation}
where $\dot{N}_{\rm raw, wing}^{\rm tot}$ is the raw total count rate of the wing, parameter $A$ links CDD($r$) with $\dot{N}_{\rm raw, wing}^{\rm tot}$ and the raw background Count Rate Density (CRD$_{\rm raw}^{\rm bkg}$, assuming the background is uniform), the area of the wing is $400\pi$\,arcsec$^2$, $\dot{N}_{\rm raw,\,ann}^{\rm COI,\,tot}(r)$ is the input value of Equation (\ref{equ:COI_point}, \ref{equ:EXT}) to calculate COI and EXT factors for the raw total count rate of each infinitely narrow annulus, denoted as ${\rm COI}_{\rm raw,\,ann}^{\rm tot}(r)$ and ${\rm EXT}_{\rm raw,\,ann}^{\rm tot}(r)$, respectively. Hence, the total count rate corrected with COI, EXT and NU is the integration of the last one in Equation (\ref{equ:dNU}):
\begin{equation}
    \dot{N}_{\rm CEN,\,wing}^{\rm tot}=2\pi\int_{\rm 15[arcsec]}^{\rm 25[arcsec]}(A\,{\rm CDD}(r)\,+\,{\rm CRD}_{\rm raw}^{\rm bkg})\,{\rm COI}_{\rm raw,\,ann}^{\rm tot}(r)\,{\rm EXT}_{\rm raw,\,ann}^{\rm tot}(r)\,rdr
    \label{equ:NU}
\end{equation}
where the subscript ``CEN" of $\dot{N}_{\rm CEN,\,wing}^{\rm tot}$ is the combination of first capitals for COI, EXT and NU, indicating corrections that have been applied. 
For the uniform case, the count rate corrected with COI and  EXT factors, denoted as $\dot{N}_{\rm CE,\,wing}^{\rm tot}$ is:
\begin{equation}
    \begin{aligned}
        \dot{N}_{\rm raw,\,wing}^{\rm COI,\,tot}&=\dot{N}_{\rm raw,\,wing}^{\rm tot}\,/\,16 \\
        \dot{N}_{\rm CE,\,wing}^{\rm tot}&={\rm COI}(\dot{N}_{\rm raw,\,wing}^{\rm COI,\,tot})\,{\rm EXT}(\dot{N}_{\rm raw,\,wing}^{\rm COI,\,tot})\,\dot{N}_{\rm raw,\,wing}^{\rm tot}
    \end{aligned}
    \label{equ:uniform_tot}
\end{equation}
For both nonuniform and uniform cases, COI and EXT corrected background count rates in the wing, denoted as $\dot{N}_{\rm CE,\,wing}^{\rm bkg}$, are same and can be calculated with Equation (\ref{equ:uniform_tot}):
\begin{equation}
    \begin{aligned}
        \dot{N}_{\rm raw,\,wing}^{\rm COI,\,bkg}&=25\pi[{\rm arcsec^2}]\times{\rm CRD}_{\rm raw}^{\rm bkg} \\
        \dot{N}_{\rm CE,\,wing}^{\rm bkg}&={\rm COI}(\dot{N}_{\rm raw,\,wing}^{\rm COI,\,bkg})\,{\rm EXT}(\dot{N}_{\rm raw,\,wing}^{\rm COI,\,bkg})\times400\pi[{\rm arcsec^2}]\,{\rm CRD}_{\rm raw}^{\rm bkg}
    \end{aligned}
    \label{equ:uniform_bkg}
\end{equation}
where $\dot{N}_{\rm raw,\,wing}^{\rm COI,\,bkg}$ is equivalent count rate used to calculate COI and EXT factors for the background count rate in the wing. A theoretical maximum correction factor for the non-uniformity (NU$_{\rm max}$) can be derived with:
\begin{equation}
    {\rm NU}_{\rm max}=\frac{\dot{N}_{\rm CEN,\,wing}^{\rm tot}\,-\,\dot{N}_{\rm CE,\,wing}^{\rm bkg}}{\dot{N}_{\rm CE,\,wing}^{\rm tot}\,-\,\dot{N}_{\rm CE,\,wing}^{\rm bkg}}
\end{equation}
Figure \ref{fig:NU_factor} presents the results of maximum NU factors (NU$_{\rm max}$) for given $\dot{N}_{\rm raw,\,wing}^{\rm tot}$ and ${\rm CRD}_{\rm raw}^{\rm bkg}$. For $\dot{N}_{\rm raw,\,wing}^{\rm tot}$ is 100(200)\,count/s, the NU$_{\rm max}$ is $\sim1.7(4.6)\%$ when the raw background count rate density is 0\,count/pixel/s. With the increasing of the background intensity, the influence of the NU become smaller for a given $\dot{N}_{\rm raw,\,wing}^{\rm tot}$.

However, NU$_{\rm max}$ is just a completely unpractical estimation for the maximum influence of the NU for 3 main reasons.
First, the coincidence loss is an area effect, hence using infinitely small regions to calculate the correction is not correct. In other words, the COI correction can not be calculated properly without the consideration of a sufficient number of pixels.
In addition, it is not hard to understand that if using a bit wider annuli to estimate the influence of the NU, (e.g., an extreme example, just divide the entire wing in to 2 annuli, one ranges from 15\,arcsec to 20\,arsec, and the other ranges from 20\,arcsec to 25\,arcsec), a smaller NU factor will be derived.
Hence, the integration method is just used to get the maximum influence of the NU, since it is not correct in practice.
Second, the area of the entire ring is $\sim5000$\,pixel$^2$, where 1\,pixel corresponds to 0.502\,arcsec. If the background count rate density is 0.006\,count/s/pixel (a median value of the V-band calibration data set, see Section \ref{subsec:generation_of_calibration_data}), the total background count rate in the wing reaches $\sim30$\,count/s that will reduce the non-uniformity of the wing, since the background count rate is almost comparable with the net count rate of the wing (i.e., total subtracts background count rate in the wing). As shown in Figure \ref{fig:NU_factor}, the curve corresponding to a background level of 0.005\,counts/s/pixel is $\sim0.4\%$ lower overall compared to the curve with zero background level.
The third reason is the CDD actually changes with the count rate and the CDD template is derived at relatively low count rate level (i.e., raw total count rate less than $\sim$60\,count/s in the wing), and for high count rate cases $>\sim80$\,count/s, the coincidence loss has stronger effect on the inner part of the wing, as a result, the inner part of the wing at high count rate level, is not as steep as that at low count rate level. Hence, the high count rate part of all curves in Figure \ref{fig:NU_factor} are expected to be slightly lower.

To summary, the non-uniformity is not included in the calibration for 2 main reasons: 1) In the valid total count rate range of PSF method ($<\sim100$\,count/s, see Section \ref{subsubsec:calibration_of_zp}), it is negligible compared with other corrections and the intrinsic dispersion of the calibration data set ($\sim20\%$, see Section \ref{subsubsec:calibration_of_zp}). 2) It is hard to get a reliable and practical NU factor since the coincidence loss is an area effect intrinsically.

\section{Calibration for individual operation modes and image bin factors}
\label{app:Individual_Calibration}
\begin{figure}[ht]
    \centering
    \includegraphics[width=\columnwidth]{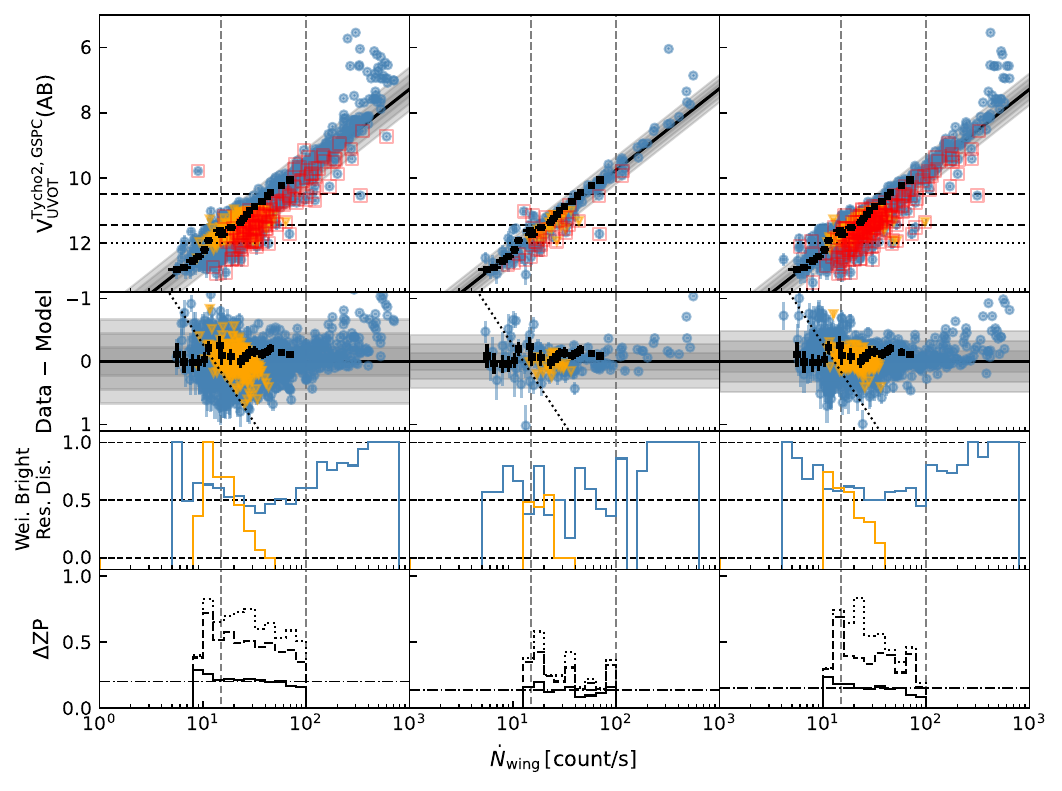}
    \caption{The V-band calibration for the PSF method in different sub data sets. From left to right, the 3 panels show calibration for event 1x1, image 1x1, and image 2x2 data sets, respectively. Please refer to the caption of Figure \ref{fig:V_all} for the meaning of all notations.} 
    \label{fig:V_3}
\end{figure}

\begin{figure}[ht]
    \centering
    \includegraphics[width=\columnwidth]{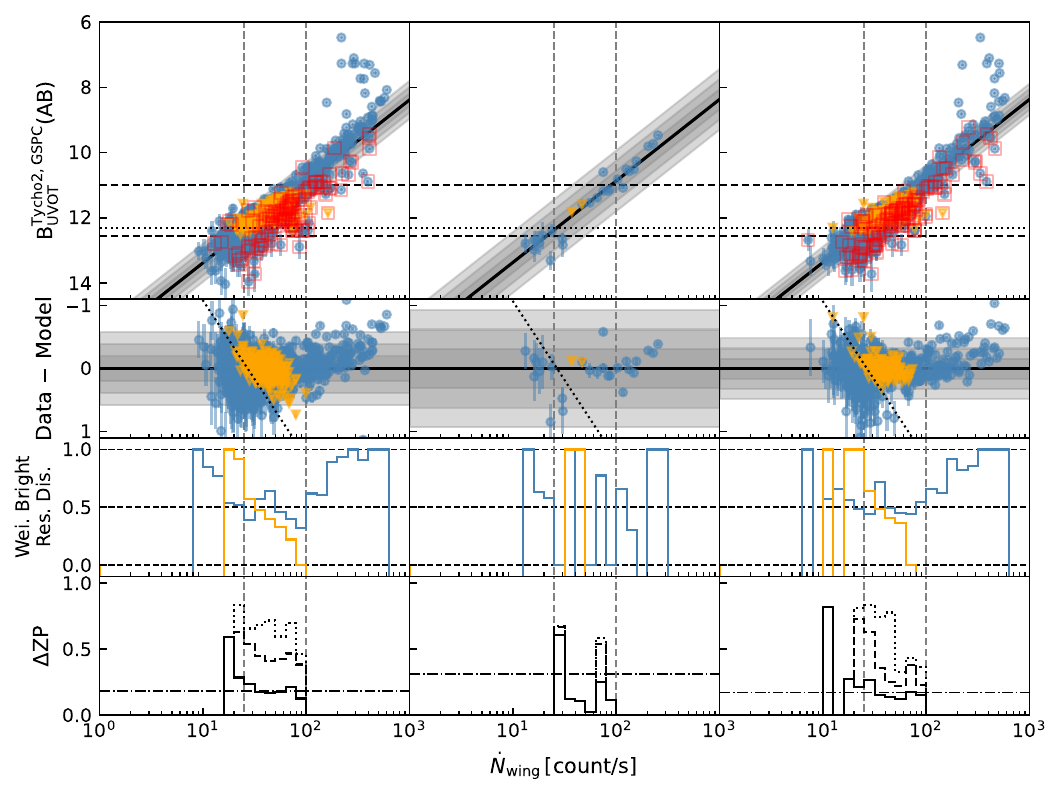}
    \caption{The B-band calibration for the PSF method in different sub data sets. From left to right, the 3 panels show calibration for event 1x1, image 1x1, and image 2x2 data sets, respectively. Please refer to the caption of Figure \ref{fig:V_all} for the meaning of all notations.} 
    \label{fig:B_3}
\end{figure}

\begin{figure}[ht]
    \centering
    \includegraphics[width=\columnwidth]{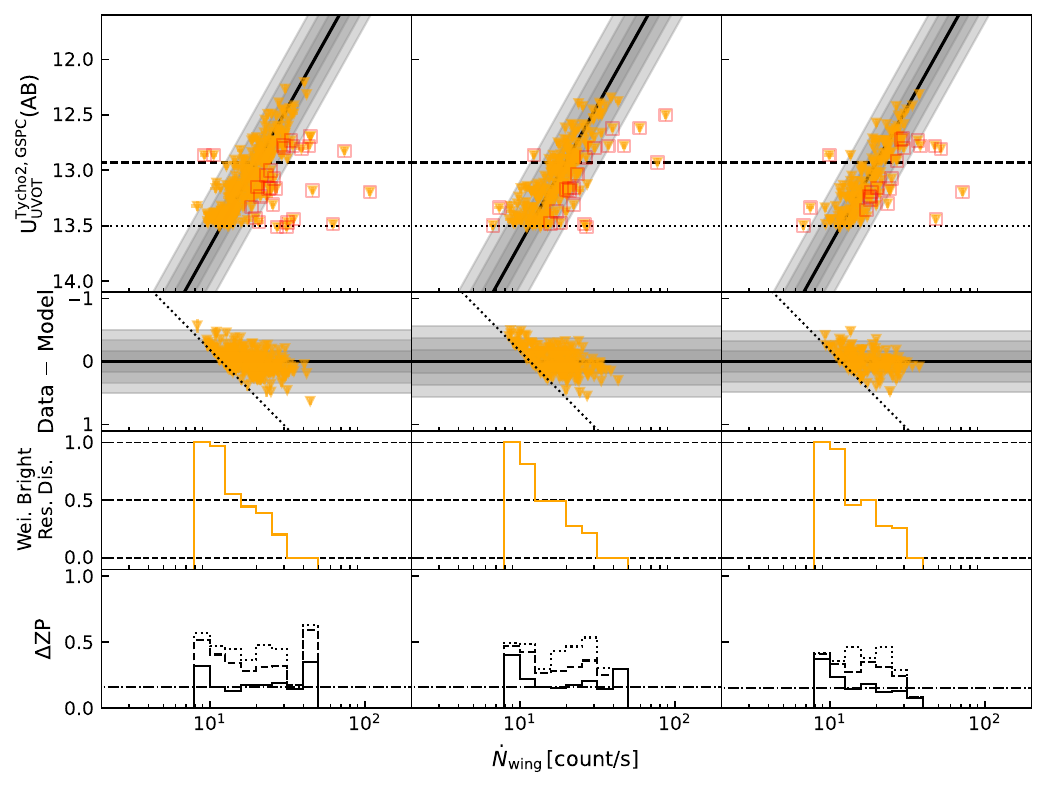}
    \caption{The U-band calibration for the PSF method in different sub data sets. From left to right, the 3 panels show calibration for event 1x1, image 1x1, and image 2x2 data sets, respectively. Please refer to the caption of Figure \ref{fig:V_all} for the meaning of all notations.} 
    \label{fig:U_3}
\end{figure}

Figure \ref{fig:V_3}-\ref{fig:U_3} show the calibration for the different operation modes and image bin factors, and Table \ref{tab:ZP_Individual} lists the calibrated zeropoints of the PSF method. There is no obvious difference between photometric properties of different operation modes and image bins for a specified bandpass, hence it is safe to conclude a zeropoint for each band. Please note it is recommended to use values in Table \ref{tab:ZP} as zeropoints of the PSF method, values in Table \ref{tab:ZP_Individual} are only listed here for any possible check.

\section{Difference between UVOT and JKC magnitude systems}
\label{app:Difference_between_UVOT_and_JKC}
\begin{figure}[ht]
    \centering
    \includegraphics[width=0.4\columnwidth]{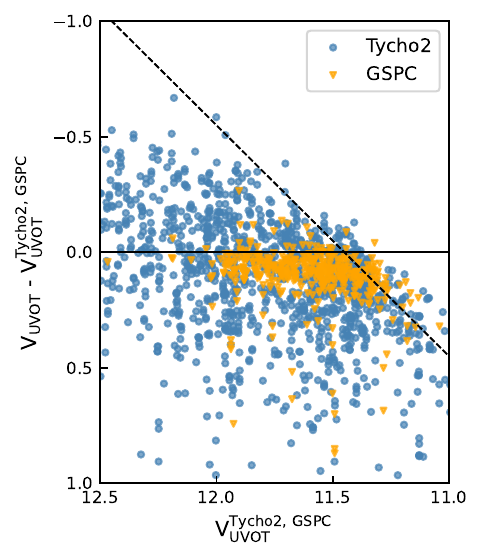}
    \includegraphics[width=0.4\columnwidth]{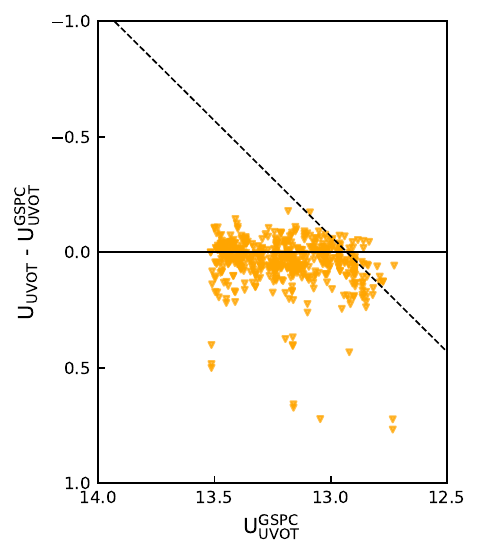}
    \caption{Difference between UVOT photometries and converted photometries from Tycho-2 and GSPC sources. The left panel is for V band, UVOT photometries are systematically fainter than converted photometires by $\sim0.08$ mag, and it can be clearly seen for GSPC data points. While due to the large scatter of Tycho-2 points, the offset is not obvious. For B band, the difference is $\sim$ 0.13 mag between UVOT photometries and converted photometries, but due to lack of GSPC points and the large scatter of Tycho-2 points, the figure for B band is not shown. However, as shown in the right panel, the difference is negligible ($\sim0.026$ mag) for U band. Dash lines in both panels represent the COI saturation limit. As for points above the COI limit, they are actually not saturated but after the LSS and the SEN correction, their corrected count rates beyond the limit. For example, the COI corrected count rate is 360\,count/s for a point source, which is smaller than the the COI limit $\sim$372, but the LSS and the SEN factors are both 1.1 for it. As a result, the corrected count rate is 435.6\,count/s, which is greater than the COI limit.} 
    \label{fig:UVOT_phot_compare}
\end{figure}

As mentioned in Section \ref{sec:Demonstration}, there are 3 possible reasons to explain why the UVOT measurements slightly fainter than RAPTOR-T measurements: 1) residual outliers in the fitted data set, 2) the intrinsic difference between RAPTOR-T V and UVOT V, 3) UVOT measurements could be slightly fainter than true values when the source is barely saturated in V band.
For the first reason, which is discussed in Section \ref{subsubsec:calibration_of_zp}, it is expected to induce a almost constant systematic bias $<\sim0.1$ mag (i.e., the difference does not likely reach a level of $\sim0.1$ mag).
For the second reason, there is indeed a rapid color evolution before $\sim400$\,s \citep[i.e., the optical spectral index evolves rapidly from $\sim0.7$ to $\sim0.4$ within $\sim100$\,s before $\sim400$\,s and keeps stable until $\sim1000$s, ][]{2009ApJ...691..723B} and such a transition from soft to hard could partly account for the $\sim0.1$ mag.
As Figure \ref{fig:VBU_Comparison} shows, the effective transmission curve of V$_{\rm JKC}$ has a red tail that extends to $\sim7000$ \AA, while the curve of V$_{\rm UVOT}$ ends at $\sim6000$ \AA.
Hence, there will be a few more red photons recorded by instruments working in V$_{\rm JKC}$ bands when observing a soft spectrum.
For the third reason, like Figure \ref{fig:UVOT_phot_compare} shows, for sources with V-band AB magnitudes falling in the range from $\sim12.00$th to $\sim11.45$th mag, their photometries measured with UVOT pipeline are slightly fainter than that converted from GSPC with a mean value of $\sim0.08$ mag.

\clearpage
\begin{table}[ht]
    \centering
    \begin{tabular}{ccccc}
        \hline
        \hline
        Filter & Valid $\dot{N}_{\rm wing}$ & $ZP^{\rm AB,\,evt1x1}_{\rm wing}\,^{a}$ & $ZP^{\rm AB,\,img1x1}_{\rm wing}\,^{a}$ & $ZP^{\rm AB,\,img2x2}_{\rm wing}\,^{a}$ \\
        & (count/s) & & & \\
        \hline
        V & 10\,-\,100 & $14.741\pm0.200$ & $14.724\pm0.139^b$ & $14.744\pm0.153$ \\
        B & 20\,-\,100 & $15.852\pm0.184$ & $15.837\pm0.311^b$ & $15.835\pm0.171$ \\
        U & 12\,-\,40 & $16.157\pm0.160$ & $16.136\pm0.162$ & $16.147\pm0.150$ \\
        \hline
    \end{tabular} \\
    $a$. ``evt1x1" means the data is recorded when UVOT works under event mode and bin factors of the image are both 1 in X and Y directions. ``img1x1" and ``img2x2" for image mode with bin factors are both 1 and 2, respectively. \\
    $b$. The number of fitted data points is too small to derive a reliable uncertainty, hence the RMS of residuals is simply presented here.
    \caption{Zeropoints of PSF method. The first column represents names of UVOT filters. The second column lists the valid range of $\dot{N}_{\rm wing}$ for each band. Next 3 columns are zeropoints of PSF method for different operation and bin modes. All values following ``$\pm$" are 1-$\sigma$ uncertainties of zeropoints, i.e., the systematic uncertainty of the PSF method. Differences between zeropoints for different operation and bin modes are very small ($<\sim0.02$ mag).}
    \label{tab:ZP_Individual}
\end{table}

\newpage
\clearpage

\newpage
\clearpage

\newpage
\clearpage

\newpage
\clearpage

\newpage
\clearpage

\end{appendix}

\end{document}